\documentclass[runningheads]{llncs}
\usepackage{amsmath}
\usepackage{comment}
\usepackage{multirow}
\usepackage[T1]{fontenc}
\usepackage[normalem]{ulem}
\useunder{\uline}{\ul}{}
\usepackage{appendix}
\usepackage{graphicx}
\usepackage{float}
\usepackage{appendix}
\usepackage{xurl}
\usepackage{algorithm2e}
\usepackage{listings}
\usepackage{graphicx}
\usepackage[dvipsnames]{xcolor}
\usepackage{pifont}
\usepackage{xcolor}

\usepackage{rotating}
\usepackage{placeins}

\usepackage{caption}
\usepackage{subcaption}
\usepackage{float}
\usepackage{multirow}
\usepackage{cite}

\begin{document}
\title{Seeing The Words: Evaluating AI-generated Biblical Art}
\titlerunning{Seeing The Words}
%
\author{Hidde Makimei\inst{1}\orcidID{0009-0004-7184-7943} \and
Shuai Wang\inst{1}\orcidID{0000-0002-1261-9930} \and
Willem van Peursen\inst{3}\orcidID{0000-0002-3142-5752}}
\authorrunning{H. Makimei et al.}
%

\institute{Department of Computer Science, Vrije Universiteit Amsterdam, the Netherlands\\
\email{hidde.n.g.makimei@student.vu.nl, shuai.wang@vu.nl}\\ \and
 Eep Talstra Centre of Bible and Computer (ETCBC), Faculty of Religion and Theology, Vrije Universiteit Amsterdam, the Netherlands\\
\email{w.t.van.peursen@vu.nl}}

%
\maketitle              
\begin{abstract}

The past years witnessed a significant amount of  Artificial Intelligence (AI) tools that can generate images from texts. This triggers the discussion of whether AI can generate accurate images using text from the Bible with respect to the corresponding biblical contexts and backgrounds. Despite some existing attempts at a small scale, little work has been done to systematically evaluate these generated images. In this work, we provide a large dataset of over 7K images using biblical text as prompts. These images were evaluated with multiple neural network-based tools on various aspects. We provide an assessment of accuracy and some analysis from the perspective of religion and aesthetics. Finally, we discuss the use of the generated images and reflect on the performance of the AI generators. 

\end{abstract}
\section{Introduction}\label{Introduction}

 The biblical text has served as a wellspring of inspiration for human creativity across various domains. Its stories, metaphors, ethical teachings, and representations of divine beings have guided and fueled the imagination of artists. Recently, there has been some primitive work on generated biblical art with Artificial Intelligence (AI). These are images generated using some recent text-to-image generators such as DALL·E 2. These generated images have been widely spread on the Internet due to their flexibility, buzzworthy effects, and little copyright concerns. More and more generated images have been used to promote churches, Bible messages, and religion-related events. In the usage of such new resources, there seems to be little reflection of the creation of these images. The elements and objects used in the generated images reflect the ``understanding'' of these generators, which may combine text and traditional interpretations, depending on the material on which the models are trained. Choosing inappropriate or inaccurate images can result in a misreading of biblical messages.  
 
\begin{figure}[!ht]

\minipage{0.32\textwidth}%
\includegraphics[width=\linewidth]{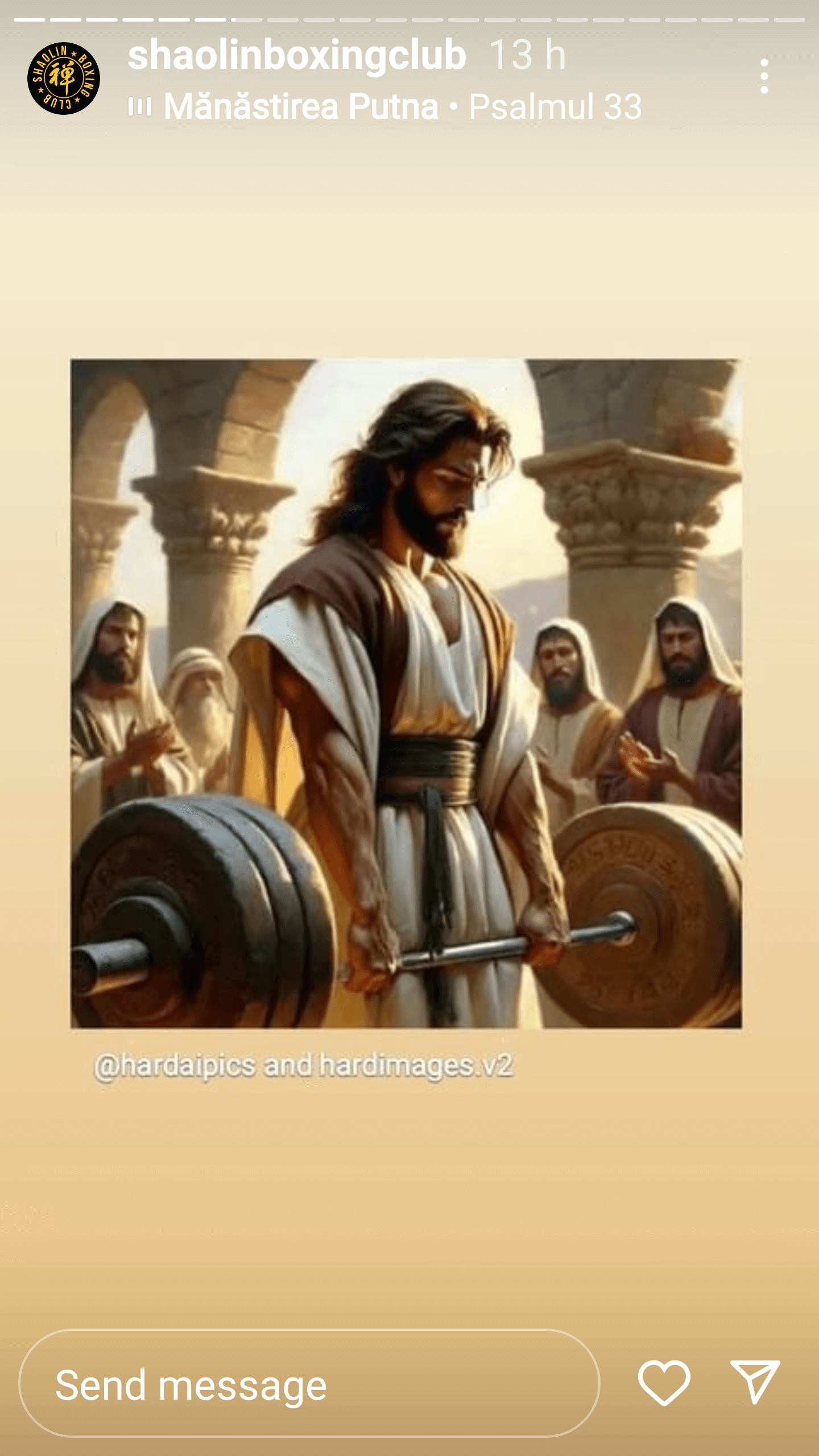}
  \caption{An advertisement  on Instagram using a generated image with Jesus lifting the weights by the Shaolin Boxing Club in Amsterdam (a screenshot with permission of use)}
  \label{fig:lift}
\endminipage
\hfill
\minipage{0.32\textwidth}%
\includegraphics[width=\linewidth]{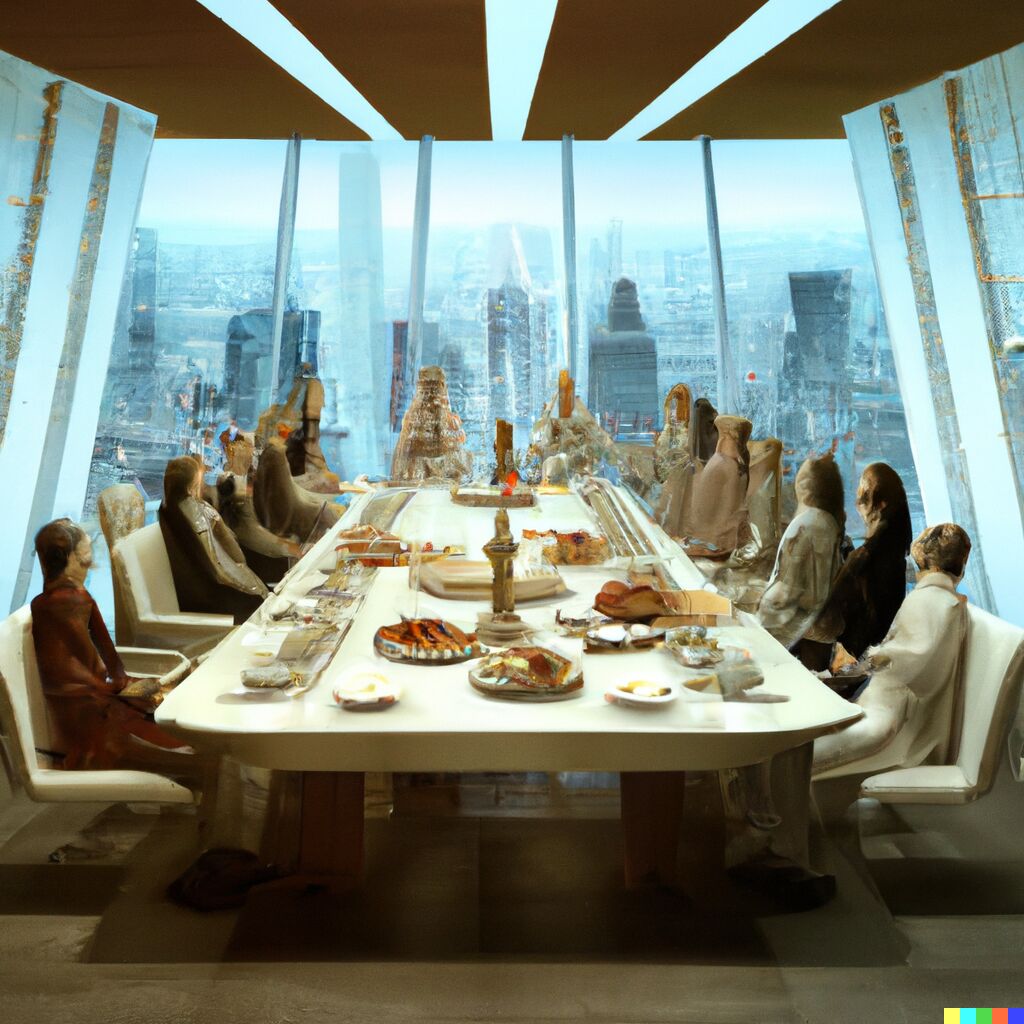}
  \caption{An image about the Last Supper generated by DALL·E, provided by OpenBible}
  \label{fig:sky}
\includegraphics[width=\linewidth]{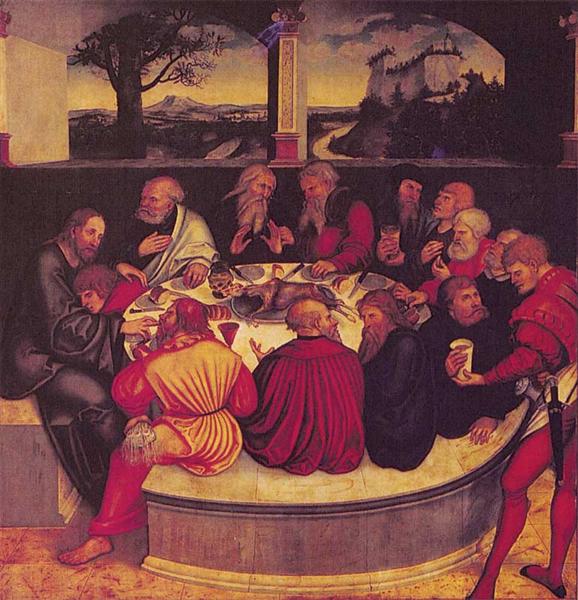}
  \caption{The Last Supper painted by Lucas Cranach the Elder in 1547}
  \label{fig:last3}
\endminipage 
\hfill
  \minipage{0.32\textwidth}
  \includegraphics[width=\linewidth]{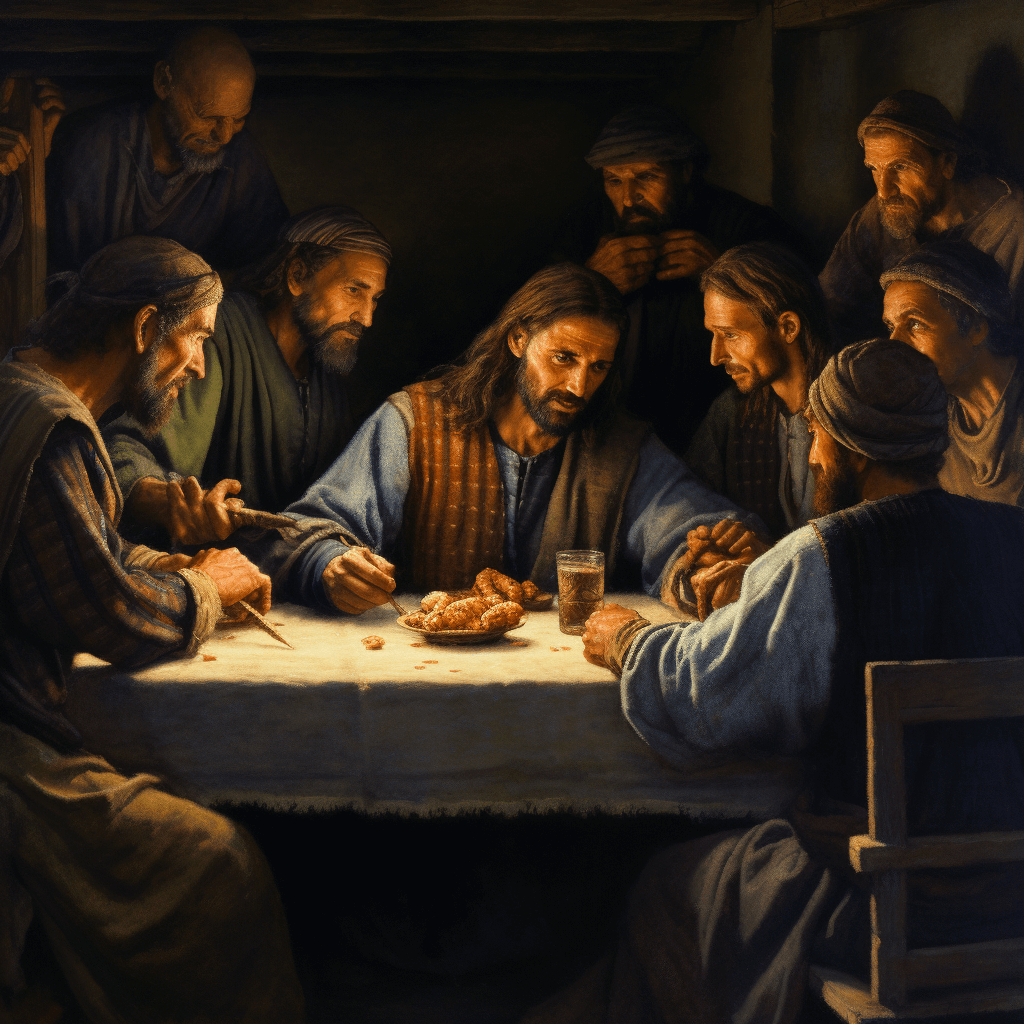}
  \caption{An image about the Last Supper generated by Midjourney in our VDD  dataset}
  \label{fig:last4}
  \includegraphics[width=\linewidth]{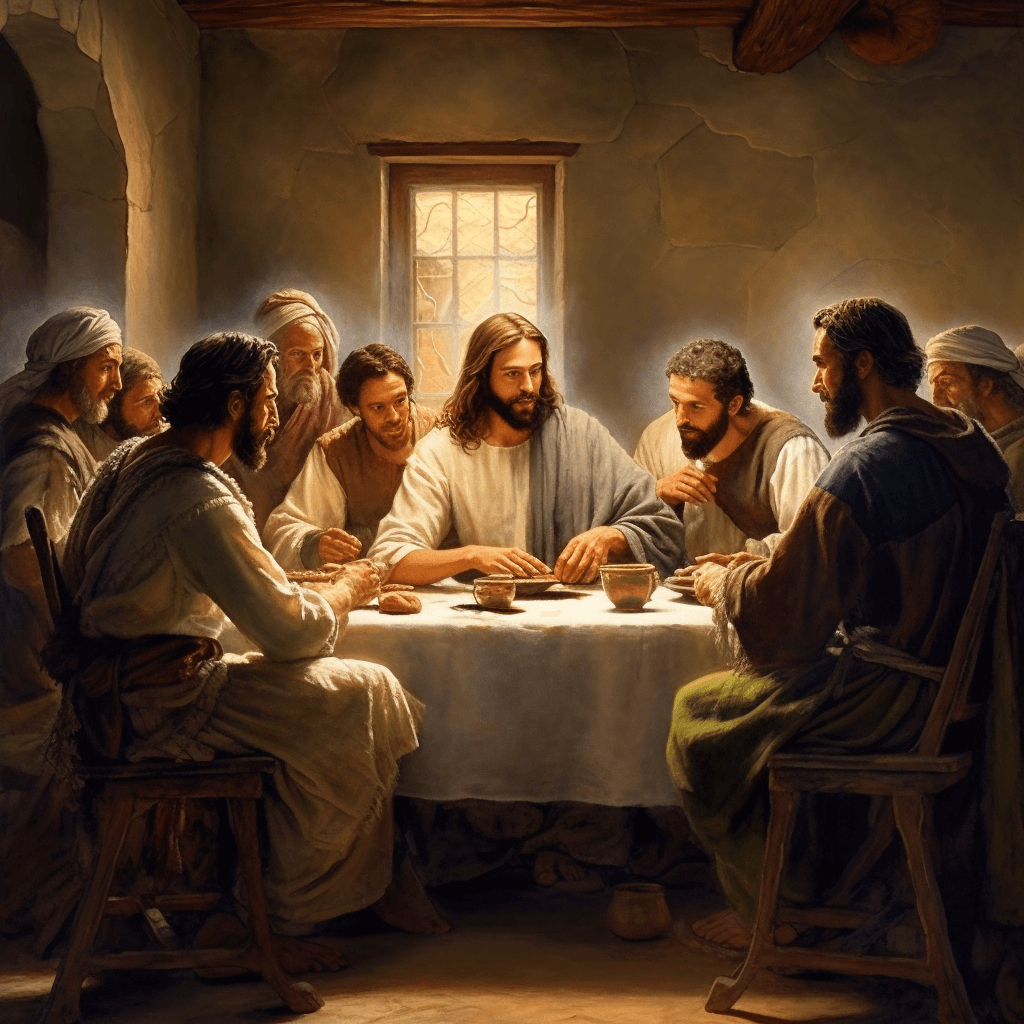}
  \caption{An image about the Last Supper generated by Midjourney in our VDD dataset}
  \label{fig:last5}
\endminipage
\end{figure}

 These images can go beyond their intended scope and combine with unexpected features. For example, Figure \ref{fig:lift} is an advertisement on social media by the Shaolin Boxing club in Amsterdam with Jesus lifting the weights. Such usage of religious motifs even raises the question whether there are any ethical or societal limits in such an "inappropriate" use of religious motifs. 
 
 To carefully study questions of accuracy, creativity, biases, the role of imagination in both traditional and AI-generated visualizations of Biblical scenes, and the ethical and societal questions involved, a large-scale assessment of AI-generated images based on biblical texts is demanded.

 To the best of our knowledge, the largest collection of generated biblical art is hosted on the OpenBible website with 1,128 images generated using DALL·E 2.\footnote{\url{https://www.openbible.info/labs/ai-bible-art/}} 
Figure \ref{fig:sky} is about the Last Supper and shows that the generator accurately captured the idea that some people sit around a table with dining setting and plates with food. The image includes surprising elements such as modern chairs, floor-to-ceiling windows, and skyscrapers in the background. The interpretation differs obviously when compared, for example, with Leonardo da Vinci's famous painting of the Last Supper.

Evaluating every element and object in these images can be hard. An accurate evaluation involves not only historical and biblical context, the aesthetics, but also careful revision by experts in the history of art, clothes, architecture, cuisine, human anatomy, etc. This is impossible when scaling to hundreds and thousands of images. In fact, including contemporary elements can also happen in human art. An example is the medieval setting (clothes, castle in the background), in Figure \ref{fig:last3}. A systematic study of these generated images requires clear standards that can be examined using an automatic approach and a reliable set of reference paintings.



This paper takes a primitive step towards systematically evaluating generated biblical images. The research questions of this paper are the following. \textbf{RQ1:} How can we systematically generate biblical images using text-to-image generators? \textbf{RQ2:} How can we evaluate the generated biblical images? For this question, we perform the evaluation in three aspects using the subquestions: \textbf{RQ2a:} What is the accuracy of persons and objects in the generated images regarding their biblical context? \textbf{RQ2b:}  How can we compare the sentimental values of the generated images? \textbf{RQ2c:} What features can be analyzed concerning religion and aesthetics?




This paper makes the following contribution.

\begin{enumerate}
    \item  We present the \textit{Visio Divina Dataset} (VDD in short)\footnote{The term `Visio Divina' is used for a practice of divine seeing by careful interaction with an image through mediation and prayer. It is a variant of the more familiar ancient tradition of the Lectio Divina, but with the reading being replaced by visual experience. For a critical evaluation of lectio divina when used as a way to avoid the academic analytical study of sacred texts, see \cite{Peursen2019}. {\color{black}The code can be found on GitHub at \url{https://github.com/ETCBC/SeeingTheWords}. The code is under continuous development, but the version corresponding to this paper is published on Zenodo with DOI: 10.5281/zenodo.14674001. Our generated images can be found on DANS SSH Data stations with DOI: \url{doi:10.17026/SS/QA271C}. Intermediate results and other supplementary materials have been archived on YODA (\url{https://yoda.vu.nl/site/}) and are available upon request.}}, a large open dataset consisting of 7,116 images from 9 text-to-image generators. 
    \item  The paper presents the results of (manual or automatic) evaluation of several features: accuracy evaluation, sentimental analysis,  religious analysis, and aesthetic analysis. 
    \item  We make the first attempt to construct a workflow and incorporate automated evaluation of AI-generated images against well-known paintings by artists referring to the same biblical text.
    \item We provide analysis from both the perspective of religion and aesthetics. 
    \item Selected images are included in an online Virtual Reality (VR) exhibition\footnote{\url{https://shuai.ai/art/seeing}}.
\end{enumerate}

This paper is organized as follows: Section \ref{sec:related} provides the related work. Section \ref{sec:data} includes details on the selection of prompts and the generation of images. Section \ref{sec:methodology} describes the methodology and presents the pipeline and evaluation. In Section \ref{sec:religious-aesthetics} we provide an analysis from the perspective of religion and aesthetics. Section \ref{sec:evaluation} presents the results of the automated evaluation. In Section \ref{sec:discussion}, we discuss the findings and limitations of the approach. Finally, Section \ref{sec:conclusion} presents the conclusion and future work. The prompts used are in Appendix \ref{appendix:prompt}. 
In Appendix \ref{sec:vr}, we provide details of our use case: an exhibition in virtual reality.

\section{Related Work}
\label{sec:related}

The BiblePics App\footnote{\url{https://biblepics.co/}} takes advantage of AI-generated images and provides visualized scenes of the Bible. As far as the authors know, the largest collection of generated art is hosted on the OpenBible website.\footnote{\url{https://www.openbible.info/labs/ai-bible-art/}} There are 1,128 images generated using DALL·E 2 including contributions from communities. Their corresponding prompts were not given, 

None of the above-mentioned datasets includes the prompt used for the generation of the images. This makes the understanding of the the generated images hard and the assessment of their context and objects impossible.

  The generated images have been argued to lack human attributes such as creativity, originality, subjectivity, emotional depth, context, cultural significance, intention, and conceptualisation \cite{chatterjee2022art, cheng2022creativity, article_Liu}. As far as the authors are aware, none of the published datasets includes an analysis of how accurately these generated images correspond to the text, nor about their aesthetics. This raises the need for a systematic assessment of images produced in this approach. A comparison of AI-generated images with well-known paintings by artists on the same topic can help understand the confounding differences between AI and humans, as well as analyze the bias of generators and guide the development of future AI-based tools. This comparison could also guide the selection of relevant images and ease manual evaluation.

\section{Data}
\label{sec:data}

\subsection{Prompt}

To unify the input of text-to-image generation, we select some representative biblical themes that have been rendered by artists with a rich amount of masterpieces. More specifically, we take five different passages as prompts. See Appendix \ref{appendix:prompt} for more details.

\begin{enumerate}
    \item  Adam and Eve's Expulsion of Paradise (Genesis 4:23-24)
    \item  The Tower of Babel (Genesis 11:1-9)
    \item Binding of Isaac (Genesis 22:9-14)
    \item  The Last Supper (Mark 14:12-25)
    \item Moses Found (Exodus 2:5-9).
\end{enumerate}

\subsection{Image generation}
\label{sec:ImageGeneration}

Since the existing work shows no systematic generation of biblical art, for a fair assessment, it is essential to provide a dataset using the same input under the same settings for all the generators. We select some state-of-the-art generators including DALL·E 2, Midjourney as well as seven different versions of Stable Diffusion. For the best performance, we used the commercial version of DALL·E 2\footnote{\url{https://openai.com/dall-e-2}} and Midjourney\footnote{\url{https://www.Midjourney.com/home/}}. For Stable Diffusion, we used some popular open-source tools: SG161222 (SG in short)\footnote{\url{https://huggingface.co/SG161222/Realistic_Vision_V1.4}}, runwayml (RW)\footnote{\url{https://huggingface.co/runwayml/stable-diffusion-v1-5}}, CompVis (CV)\footnote{\url{https://huggingface.co/CompVis/stable-diffusion-v1-4}}, stabilityai (SAI)\footnote{\url{https://huggingface.co/stabilityai/stable-diffusion-2-1}}, prompthero (PH)\footnote{\url{https://huggingface.co/prompthero/openjourney-v4}}, nitrosocke (NS)\footnote{\url{https://huggingface.co/nitrosocke/Ghibli-Diffusion}}, and dreamlike-art (DA)\footnote{\url{https://huggingface.co/dreamlike-art/dreamlike-photoreal-2.0}}. Since Midjourney lacks an API, we customized a bot that takes over the computer and interacts with the Midjourney bot for the automatic collection of generated images. For DALL·E 2, we used its API. For all variants of Stable Diffusion, we generated the images on the Google Colab cloud server that uses the A100 GPU. All the generators were accessed in the week of 19th of June, 2023. All the images are associated with a unique code for easy reference.

The images were produced through an automated process where the prompts were fed repeatedly into the generators with a summary in Table \ref{tab:dataset}. For DALL·E 2, the size of the prompt exceeded the character limit. Thus, prompts 2 and 4 were reduced by using NLTK Library\footnote{\url{https://www.nltk.org/}} with stopping words and punctuation removed.

\begin{table}[!ht]
\centering
\caption{A summary of AI generators and their generated images}
\begin{tabular}{l|l|l|lllllll|l}
\hline
\multirow{2}{*}{} & \multirow{2}{*}{DALL·E} & \multirow{2}{*}{Midjourney} & \multicolumn{7}{|l|}{Stable Diffusion} &  \multirow{2}{*}{Sum (VDD)} \\ \cline{4-10}
 &  &  & RW & CV & SAI & PH & SG & NS & DA  &\\ \hline
Version &V1 beta & V5.1 &V1.5 & V1.4 &   V2.1 & V1.1 & V1.4 & V1.1 & V2.0  & \\ 
\#images & 500 & 616 & 1K & 1K & 1K &  1K & 1K & 500 & 500 & 7,116 \\ \hline
\end{tabular}

\label{tab:dataset}
\end{table}



\begin{table}[!ht]
\caption{{\color{black}An overview of the paintings chosen}}
\label{tab:paintings}
\centering
\begin{tabular}{l|p{4.5cm}|p{4.5cm}|p{1.7cm}}
\textbf{Prompt} & \centering\textbf{Painter} & \centering \textbf{Title} & \textbf{Estimated Year of Painting} \\ \cline{1-4}
\multirow{5}{*}{prompt 1} & Michelangelo (1508–1512) & The expulsion from paradise & 1508-1512 \\
 &  Jan Brueghel II (1601-1678) & The expulsion from paradise & 1650 \\
 & Benjamin West (1738–1820) & The expulsion of Adam and Eve from Paradise & 1791\\
 & Izaak van Oosten (1613-1661) &  The expulsion of Adam and Eve from Paradise by the angel & 1628-1661 \\
 & Cornelis van Poelenburg (1594–1667) & The Expulsion from Paradise  & 1646 \\ \hline
\multirow{5}{*}{prompt 2} & Lucas van Valckenborch (1535-1597) & Tower of Babel & 1594  \\
 & Pieter Breugel (1525/30–1569) &  Tower of Babel  & 1563 \\
 &Abel Grimmer  & The tower of Babel &  1604\\
 & Hendrick van Cleve III (1525-1590) & Tower of Babel  & 1570 \\
 & Frederik van Valckenborch (1566 -1623) & The construction of the Tower of Babel & 1600\\ \hline
\multirow{5}{*}{prompt 3} & Rembrandt van Rijn (1606–1669) &  The angel prevents the sacrifice of Isaac & 1635 \\
 & Titian (a.k.a. Tiziano Vecelli, 1488/1490-1576)  & Abraham and Isaac & 1542-1544 \\ 
  & Carvaggio (1571-1610) & Sacrifice of Isaac & 1598-1603 \\ 
   & Lucas Gassel (1490–1568) &  
The sacrifice of Isaac  &1539 \\ 
  & Bartolomeo Cavarozzi (1587–1625)  &Sacrifice of Isaac &  1598 \\ \hline
 
\multirow{5}{*}{prompt 4} & Leonardo da Vinci (1452–1519) & The last supper & 1495-98 \\
&Hans Holbein de Jonge (1497–1543) &  The last supper & 1527 \\ 
&Tintoretto (a.k.a. Jacopo Robusti, 1518–1594) & The last supper &1592-94  \\ 
&Peter Paul Rubens (1577-1640) &   The last supper &1630-1631  \\ 
& Juan de Juanes (1523–1579) & The last supper & 1560 \\ \hline

\multirow{5}{*}{prompt 5} & Rembrandt van Rijn (1606–1669) & Moses found & 1635 \\
&Jan Kosten (17th century) & {\color{black}The finding of Moses} & 17th century\\
& Toussaint Gelton (1630–1680) & Moses is found by Pharaoh's daughter & 1645-1680 \\
& Paolo Veronse (1523-1580) & The finding of Moses  & 1570-1575 \\  
& Bartholomeus Breenbergh (1580-1640) & The finding of the infant Moses by Pharaoh's daughter & 1622 
\end{tabular}
\end{table}

\subsection{Artwork}
\label{sec:artwork}
The biblical artwork chosen to compare to the AI-generated images are paintings from the Renaissance and Baroque periods {\color{black} (Table \ref{tab:paintings}).} Choosing a time period narrows down the sample group of biblical art to more similar like-minded artists. The Renaissance shows the emergence of a naturalistic style (compare, e.g., the interest that painters developed in anatomy, proportions and perspective). This was further developed in the Baroque, which is well-known for its use of contrast, movement, exuberant detail, deep colour, grandeur, and surprise to achieve a sense of awe, in other words, to express sentiment. These features render paintings from these style periods good candidates for automatic analysis (e.g., object or sentiment recognition).

{\color{black} The selection of the paintings took into consideration the visibility of the characters and the accessibility of clear scans of the original painting to reduce the mistakes of the chosen machine learning models while evaluating them. These paintings form the base data for evaluation in the steps to be described in the next section.}



\section{Methodology}
\label{sec:methodology}

\begin{figure}[!ht]
    \centering
\includegraphics[width=12.5cm]{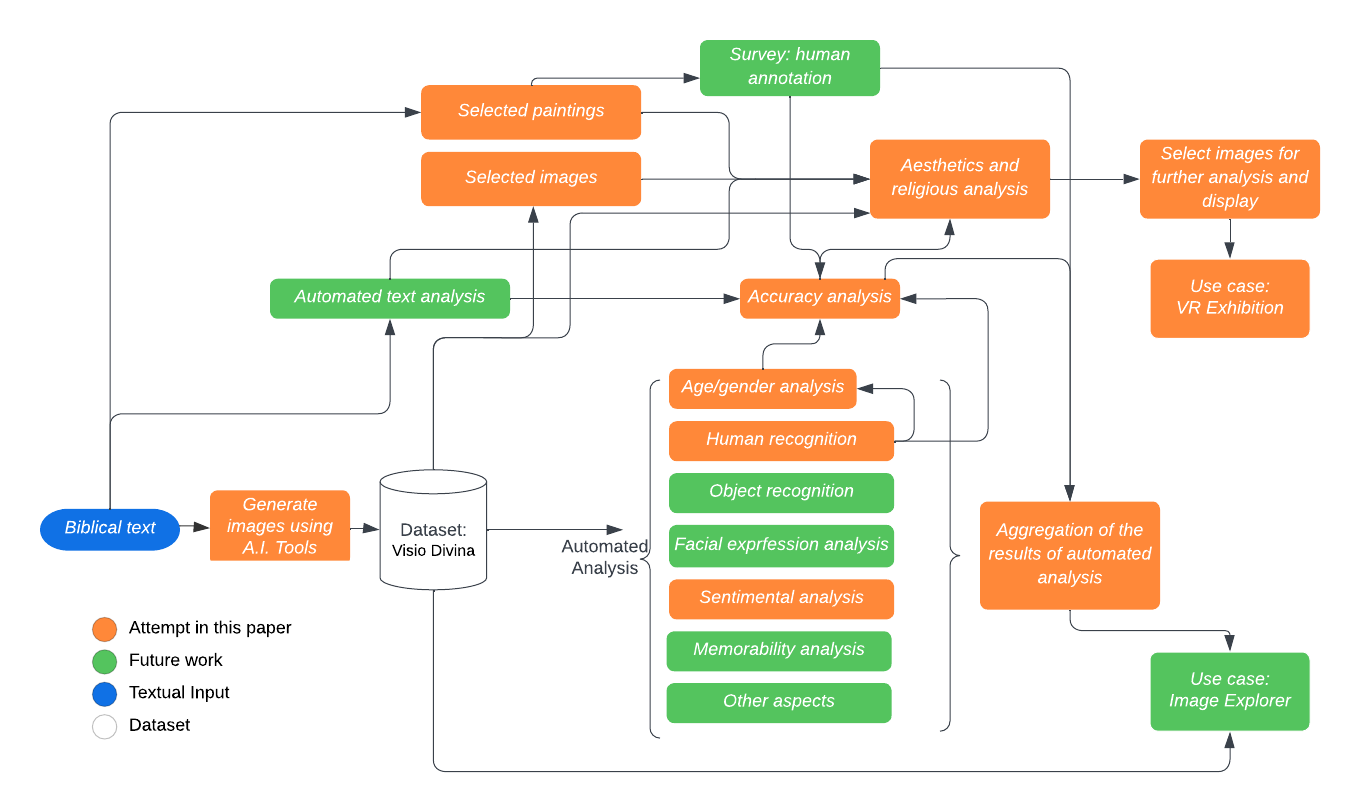}
    \caption{Workflow of image generation, evaluation against selected paintings and their use cases}
    \label{fig:workflow}
\end{figure}

{\color{black}Figure \ref{fig:workflow} is workflow that visualizes the steps taken in the study (in orange) and some related future work that has not been implemented in this work (in green). We take selected biblical text as input. Some text-to-image A.I. tools were used for the generation of images. These generated images form a dataset Visio Divina. Generated images were manually evaluated by comparing against selected paintings in two ways. We do so by comparing the selected paintings against all the generated images on the same topic (for an overall impression of the style, accuracy, sentiment, and other aspects of aesthetics), as well as some selected generated images of artistic interest for a more detailed comparison. Manual evaluation could also take advantage of automated text analysis for the evaluation of accuracy (e.g. regarding objects of religious importance and historical accuracy, etc.).  Moreover, its aesthetics as well as some other religious analysis could only be manually analyzed by taking advantage of experts' knowledge. Some images of artistic interest can be selected for discussion and exhibition. Future manual evaluation includes comparing human annotations for selected paintings and generated images.}

For automated analysis, we focus on two aspects: the people and sentiment. As for the people in the generated images, we take advantage of state-of-the-art neural network models for the detection and evaluation of the number of people. We provide some analysis of their ages and gender. 
{\color{black}Excluded from automated analysis in this paper due to the capacity of the authors are the weather, objects, memorability analysis, and facial expressions of the people. Other excluded aspects include the style of clothes, supernatural beings (e.g. angels and ghosts), biblical and historical implications, etc. They are either difficult to evaluate or not directly related to the biblical context. The automated analysis of text is left for future work. Although there are models for predicting the genre \cite{agarwal2015genre} and the style \cite{shamir2012computer}, they are not the major aspects we study and can be more about artistic style than} the biblical context in this study. Small objects such as a knife and an apple are mentioned in the prompts and could be added to the workflow and could be included in future work. {\color{black} Although it is almost impossible to accurately manually annotate every generated image, it could be possible for a relatively small number of selected paintings.}


\subsection{Religious Analysis and Aesthetics}
\label{sec:methodology-religious-aesthetics}
 As indicated in section \ref{sec:artwork}, we compare the AI-generated images with human artwork based on the same passages that served as prompts for the generated artwork. The selected paintings come from the the Renaissance and Baroque periods because of their naturalistic style. The underlying assumption is that these paintings render the biblical stories in a naturalistic or realistic way and therefore can be used to evaluate the accuracy of the AI-generated images.
 
 The recognition of elements such as human beings, gender, age, objects, sentiment, and landscape in these paintings was done manually with a survey among about fifty participants. Just like the automatic recognition tools, the responses showed variation in the identification of humans, objects and sentiments. We used measures similar to those described in Section \ref{sec:measures} to analyse and average the responses. This concerned not only the different interpretations by the human respondents of the same painting, but also the differences between the various paintings. The outcome served as a baseline with which the AI-generated images could be compared.

 The classic Renaissance and Baroque paintings provided a powerful yardstick for comparison, but some caveats should be mentioned here. First, the interpretation of these paintings is not as unequivocal as it seems. Human interpreters differed about the age or gender of people in the paintings and even more about sentiment.
 
 Secondly, it should be noted that we use here “naturalistic” as referring to  the way in which the world, landscapes, or human bodies were depicted. The study of perspective, proportions or perspective found in the Renaissance found its way into art and the further development of the effects of light or the expression of emotion in the Baroque. This does not mean, however, that the pictures were realistic in the sense that, for example, biblical scenes reflect people, clothes or landscapes of ancient Israel. It is clear that those paintings of biblical scenes, just as any painting, looks at those scenes through the lens of the artist’s time. The way in which this happens is a research object in itself (cf. above, Section \ref{Introduction} on the medieval setting of {\color{black} Figure \ref{fig:last3}).} For our purposes a certain faithfulness to the stories (e.g. in paintings of Moses being found by Pharao's daughter in the Nile, a river, a princess, and a baby are represented) helps to use those paintings as a source for training and comparison for the AI-generated pictures.

Thirdly, the AI-generated images are not completely independent from the human art works selected. Some of these paintings, like Leonardo da Vinci's Last Supper or Pieter Bruegel's Tower of Babel belong to the most famous works of art history and have shaped the Western imagination of these scenes and this affects, as we shall see (cf. especially Section \ref{sec:StableDiffusion}), to some extent also the AI-generated images. 

 Accordingly, the selected paintings are a useful benchmark for evaluating the AI-generated images, but we use them carefully, taking into account that these very pictures may have caused biases in the generated images and avoiding the assumption that these paintings represent an unequivocal, correct representation of the text against which the generated images should be assessed.

\subsection{Analysis of Human Beings}
\label{sec:human-analysis}

To answer our research question RQ2a, for the analysis of human beings in the images and paintings, we focus on three aspects: the number of humans as well as their age and gender.

\subsubsection*{Human Recognition with Detectron2} 
\label{sec:human-recognition}
Detectron2 \cite{wu2019detectron2} is a Mask R-CNN (Region-based Convolutional Neural Networks) taking advantage of ResNet-50 \cite{he2016deep} and FPN (Feature Pyramid Network) \cite{lin2017feature}. It uses Mask R-CNN and extends the Faster R-CNN model by masking in order to achieve pixel-wise segmentation. Its Mask R-CNN includes  four layers of 3×3 convultion applied to a 14×14 input feature map, whose output passes through a deconvelution layer which gets transformed using a 2×2 kernel. The neural network ends with a 1x1 convolution network that predicts the mask logits. This model is used for mapping the segmentation after training on the COCO dataset \cite{lin2014microsoft} with 8 categories and the Cityscape dataset \cite{cordts2016cityscapes}. We only identify the labels corresponding to humans identified for each given image and use those with a confidence score of 0.8 or higher. The outputs are some bounding boxes for each person detected, which are used to count the number of human characters detected in the images in this study. These bounding boxes are then used for the estimation of age and gender in the next step.

\subsubsection*{Age and Gender Estimation}
\label{sec:age-gender}

For age and gender estimation, a custom CNN \cite{LH:CVPRw15:age} was developed by Gil Levi et al. based on LeNet-5, whose main architecture consists of three convolutional layers and two fully connected layers. Each layer of the CNN is followed by ReLu and normalization before being passed on to the next. Finally, the fully connected layers are mapped to the final phase that can classify the age and gender respectively. For this model, the ImageNet dataset\footnote{\url{https://www.image-net.org/}} was used for training. The network produces an age prediction in the form of a range with a minimum and a maximum. For this pilot study, the estimated age is taken as the mean of the minimum and the maximum. We take the predicted gender in a simplified setting: male or female. Non-binary cases are beyond the scope of this work but could be addressed in future work.

\subsection{Sentimental Classification}
\label{sec:sentimental}

To answer the research question RQ2b, for sentimental recognition, we use a model introduced by Victor Campos et al.\cite{CAMPOS201715} Using an AlexNet-styled network \cite{krizhevsky2012imagenet} composed of five convolutional layers and three fully-connected layers. The model passes the pixel value through the CNN to obtain an overall sentimental value of the image. It takes a dataset retrieved from Twitter with 1,269  tweets and their corresponding images with sentiment labels obtained by crowd-sourcing as training data \cite{you2015robust}. One observation is that it tends to map brighter pixels to more positive sentiment. The resulting sentimental value is in an interval between 0-1 (1 for positive). The model can be altered into a fully convolutional network with no additional training need. This produces kernel $8 \times 8$ predictions maps of the image giving 64 patches of the image with its own sentimental value. Given that the resulting sentimental values would differ if the two networks differ, we evaluate the result corresponding to two different settings.

\begin{table}[]
\caption{Models used and their training datasets}
\label{tab:models}
\footnotesize
\begin{tabular}{p{2.4cm}|p{1.4cm}|p{4cm}|p{4cm}}
 & \textbf{Aspect} & \textbf{Core Models} & \textbf{Training Dataset} \\ \hline
\multirow{3}{*}{Human Analysis} & Number of people & Detectron2 (Mask R-CNN and ResNet-50 \cite{he2016deep}) \cite{wu2019detectron2} & COCO\cite{lin2014microsoft}, Cityscape\cite{cordts2016cityscapes} \\ \cline{2-4}
 & Age & LeNet-5 \cite{LH:CVPRw15:age} & ImageNet \cite{krizhevsky2012imagenet} \\ \cline{2-4}
 & Gender & LeNet-5 \cite{LH:CVPRw15:age} & ImageNet \\ \hline
\multicolumn{2}{c|}{Sentimental Analysis }  & AlexNet \cite{krizhevsky2012imagenet}  & A dataset of tweets and images retrieved from Twitter with crowd-sourced sentimental labels \cite{you2015robust} 
\end{tabular}
\end{table}

\section{Religious Analysis and Aesthetics}
\label{sec:religious-aesthetics}

While the pipeline and evaluation compare the images to Renaissance paintings, they do not address the accuracy of religious context and aesthetic features in the images. In order to answer SRQ2c, we manually examine the generated images and study the importance of religious accuracy and aesthetic quality. This research question is vital given the interdisciplinary nature of the task. 

\subsection{DALL·E}
\label{sec:method-DALLE}

\begin{figure}[!ht]
\minipage{0.32\textwidth}
  \includegraphics[width=\linewidth]{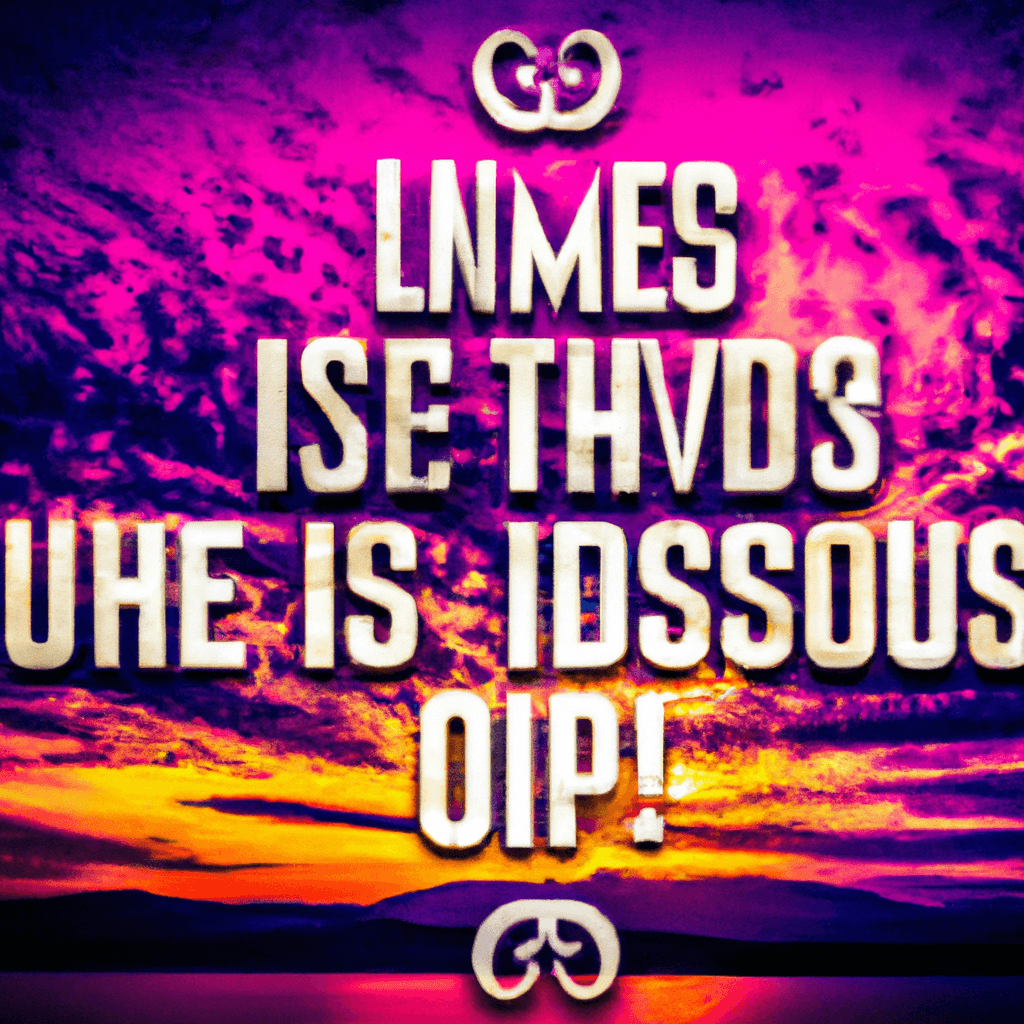}
  \caption{An image generated by DALL·E with incomprehensible text over some scene of nature using prompt 4 (the Last Supper)}
  \label{blah_1}
\endminipage\hfill
\minipage{0.32\textwidth}
  \includegraphics[width=\linewidth]{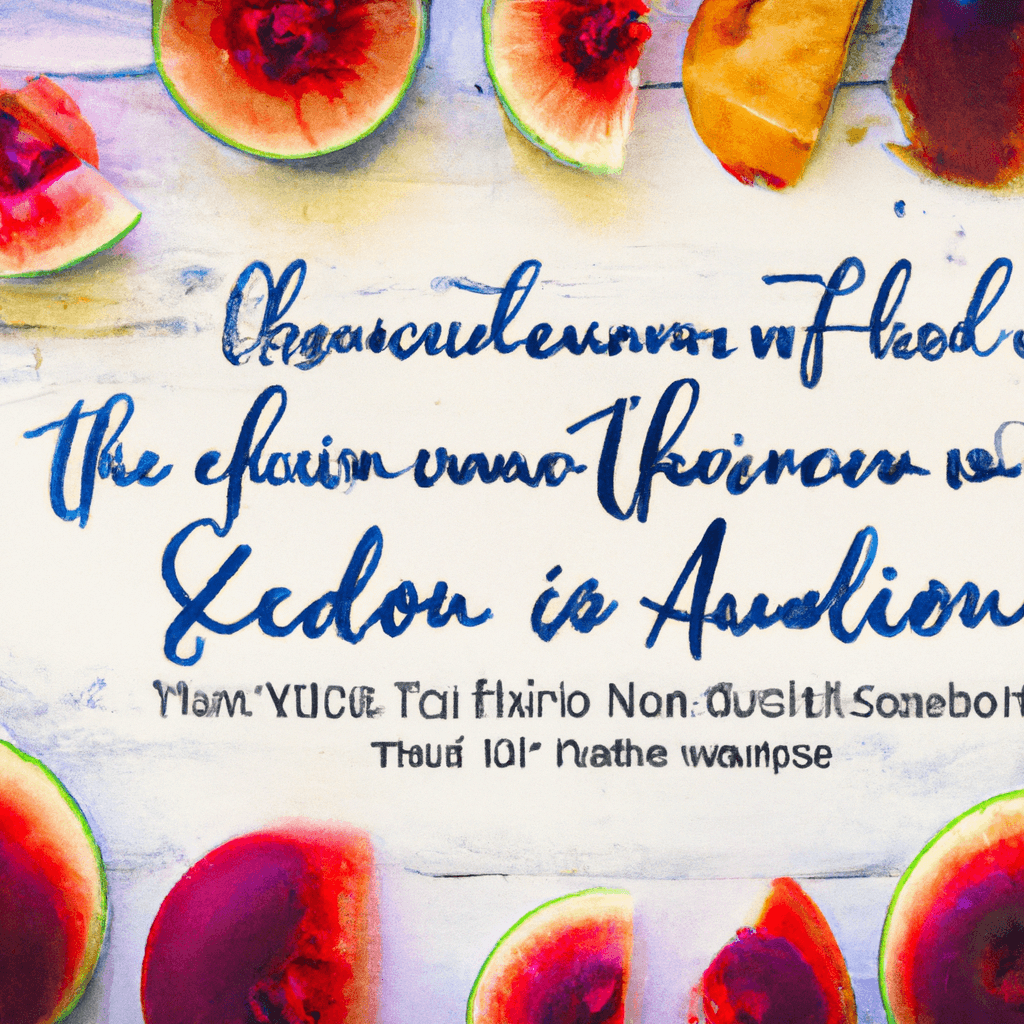}
  \caption{An image generated by DALL·E with incomprehensible text using prompt 4 (the Last Supper)}
  \label{blah_2}
\endminipage\hfill
\minipage{0.32\textwidth}%
  \includegraphics[width=\linewidth]{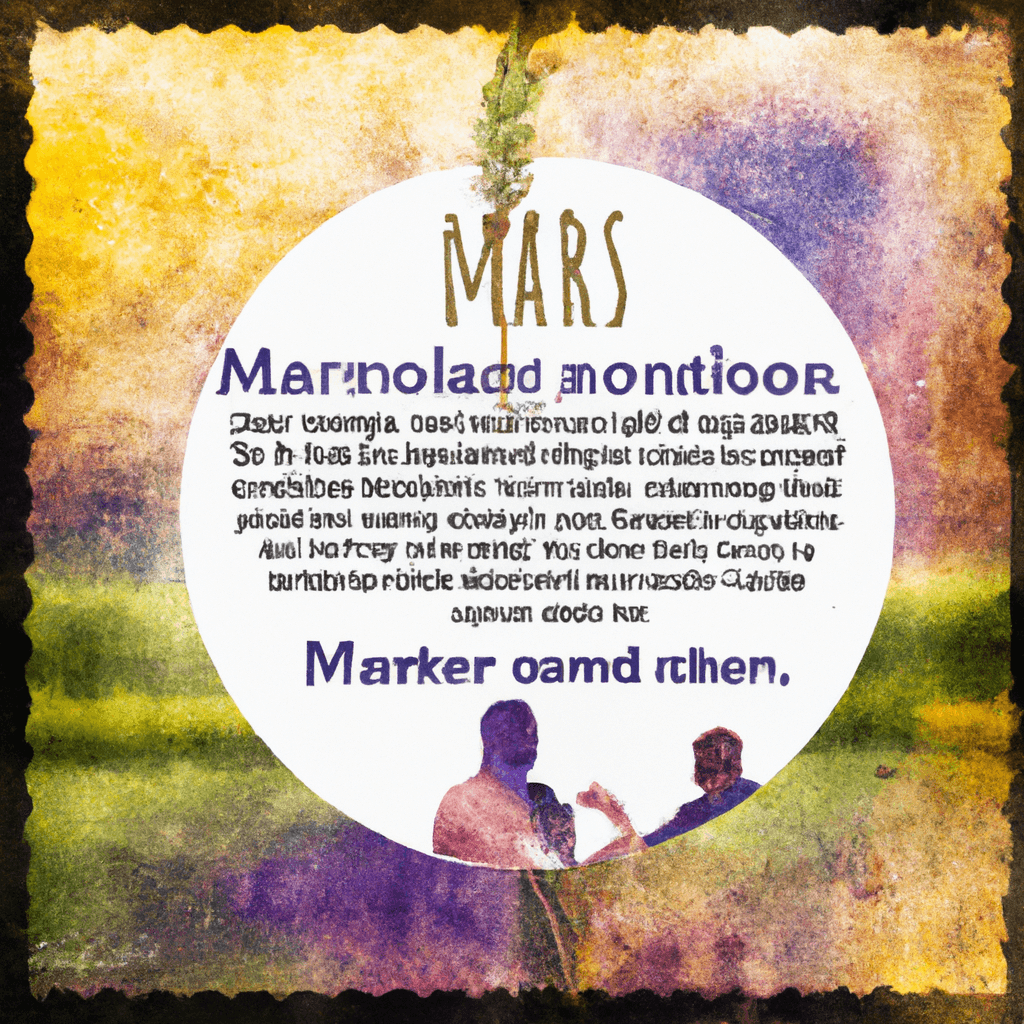}
  \caption{An image generated by DALL·E with incomprehensible text using prompt 4 (the Last Supper)}
  \label{blah_3}
\endminipage

\end{figure}

Our manual examination shows that DALL·E 2 has produced the least accurate text images related to the prompt. The generated images usually do not incorporate the elements or features described in the text. For example, in prompt 4, the generator produces mainly incomprehensible text as seen in Figures \ref{blah_1}, \ref{blah_2} and \ref{blah_3}. These images have no link to the text and do not contain any reference or attributes to the Last Supper.

\begin{figure}[h!]
\minipage{0.32\textwidth}
  \includegraphics[width=\linewidth]{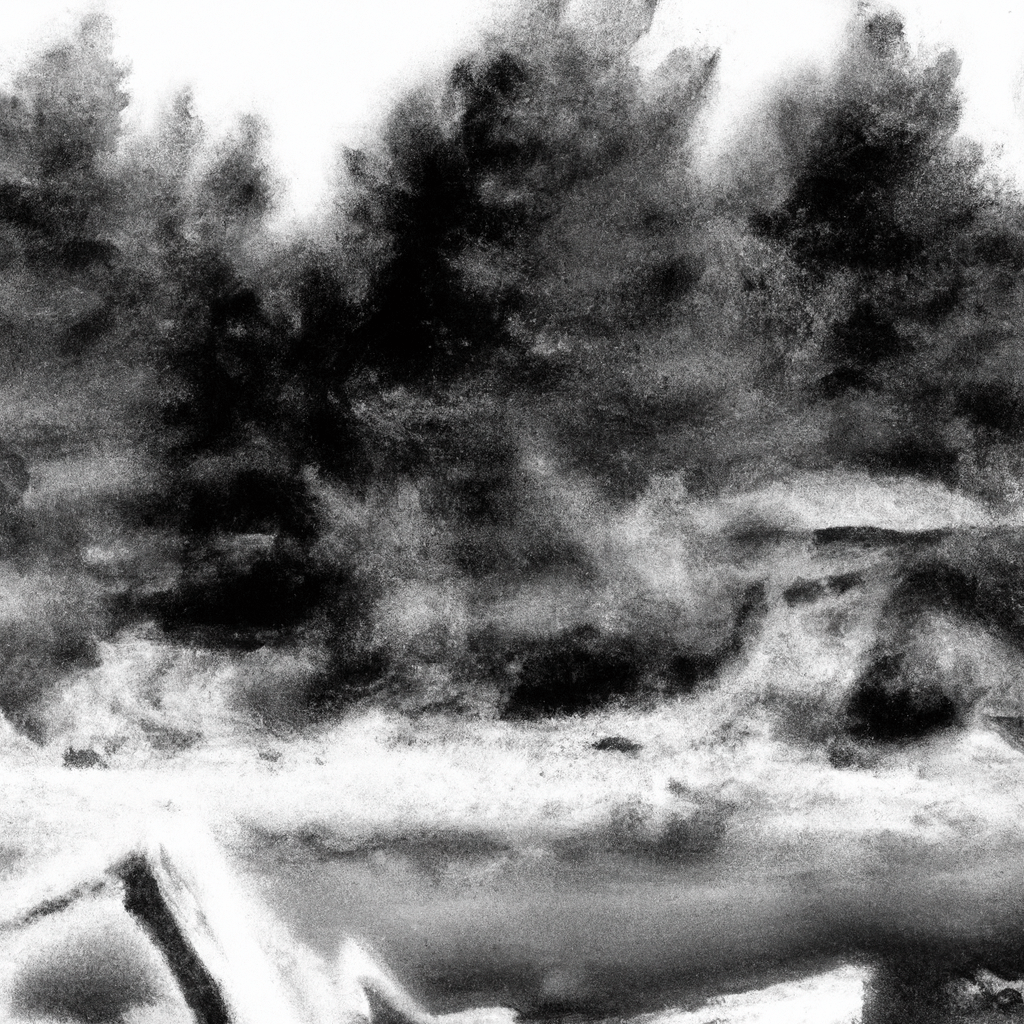}
  \caption{An image generated by DALL·E using prompt 2 (constructing the Babel Tower) with forest}
  \label{arrid_1}
\endminipage\hfill
\minipage{0.32\textwidth}
  \includegraphics[width=\linewidth]{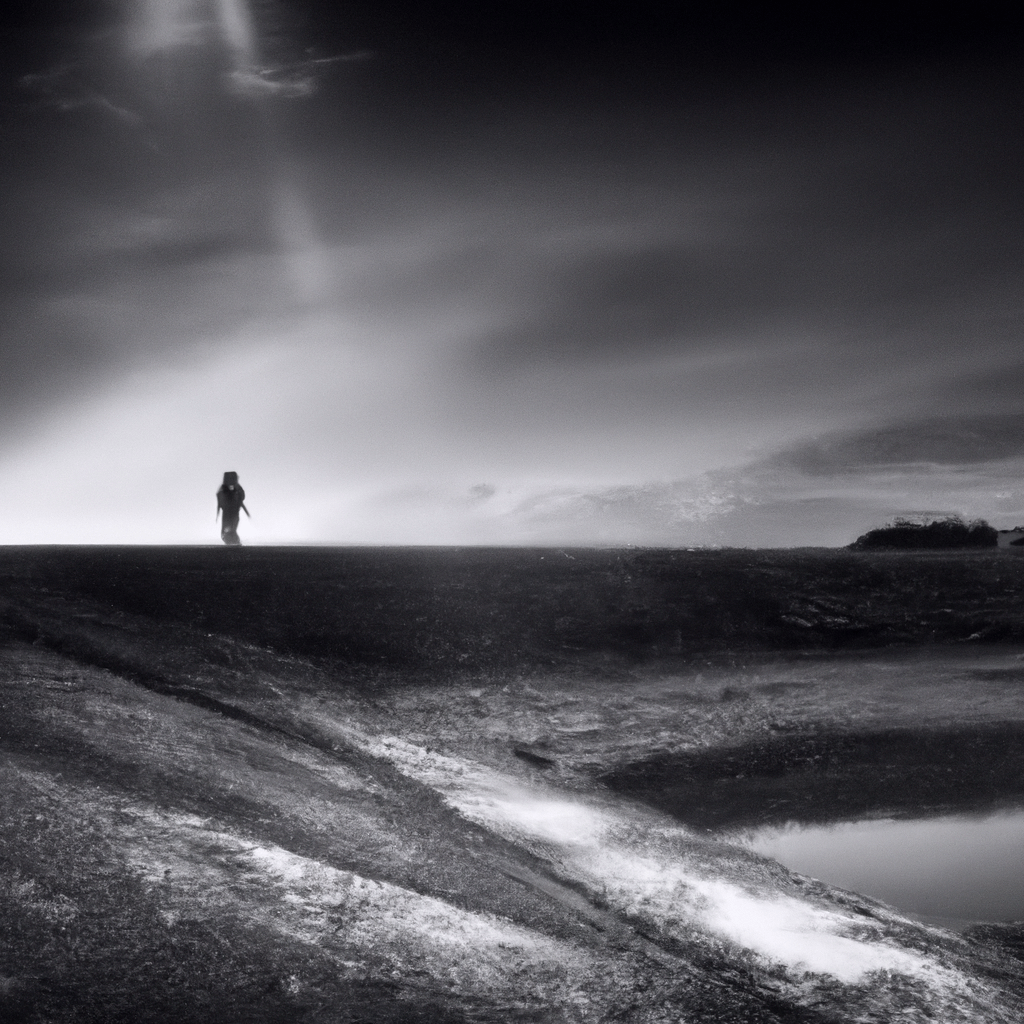}
  \caption{An image generated by DALL·E using prompt 2 (constructing the Babel Tower) with landscape}
  \label{arrid_2}
\endminipage\hfill
\minipage{0.32\textwidth}%
  \includegraphics[width=\linewidth]{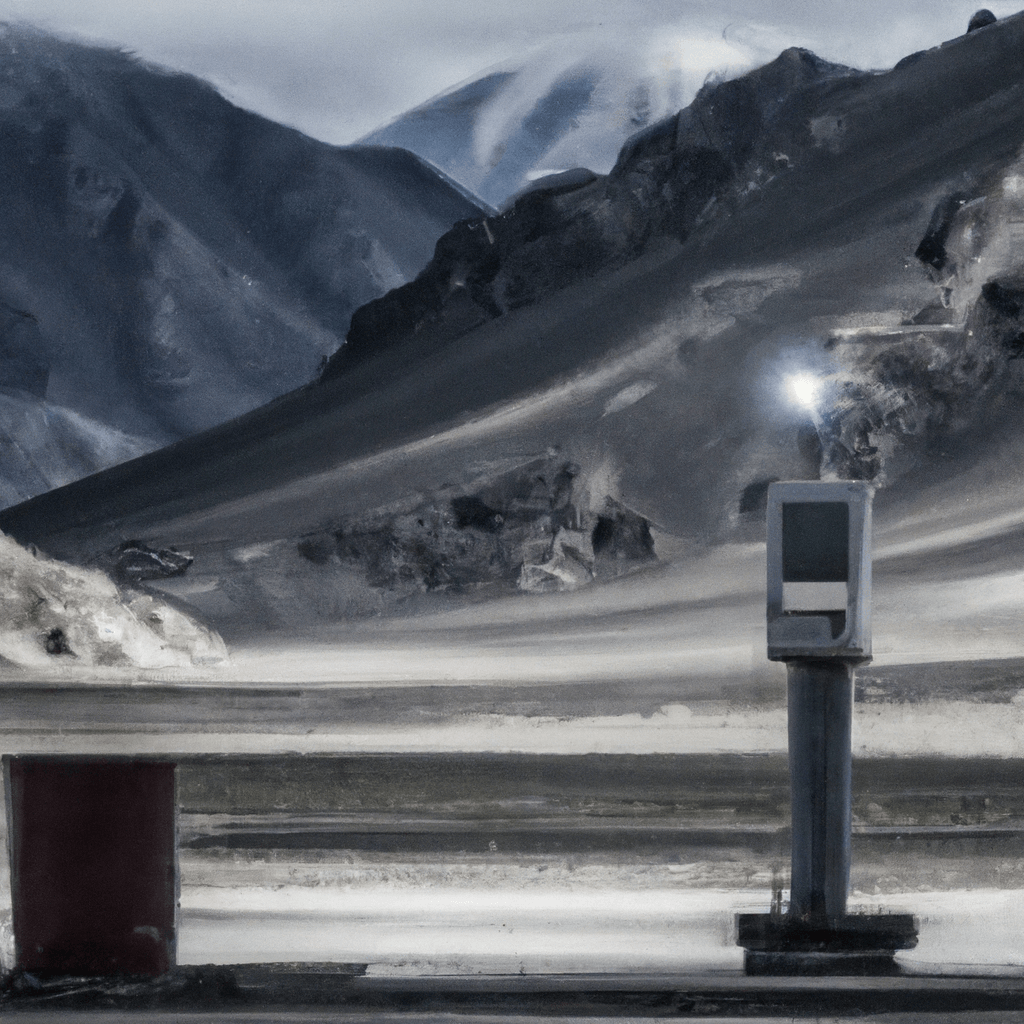}
  \caption{An image generated by DALL·E using prompt 2 (constructing the Babel Tower) with landscape}
  \label{arrid_3}
\endminipage

\end{figure}

 Prompt 1 and prompt 3 have biblical references with images incorporating angel wings, halos and medieval glass that is commonly seen in churches. This shows some biblical reference in the images for some prompts, but it still lacks in connecting the image to the actual story referred to in the prompt. On the other hand, prompt 2 mainly generates arid landscapes or forests, which may relate to phrases from the prompt such as ``the whole earth'', ``a plain'', ``the face of the whole earth'', and ``abroad over the face of the whole earth''. The images from prompt 2 do not show any accuracy to the story of the Tower of Babel, only the incomprehensible landscape, which has no relationship to the prompt as seen in Figures \ref{arrid_1}, \ref{arrid_2} and \ref{arrid_3}. This shows that DALL·E failed to comprehend the biblical context. 
 
\begin{figure}[!ht]
\minipage{0.32\textwidth}
  \includegraphics[width=\linewidth]{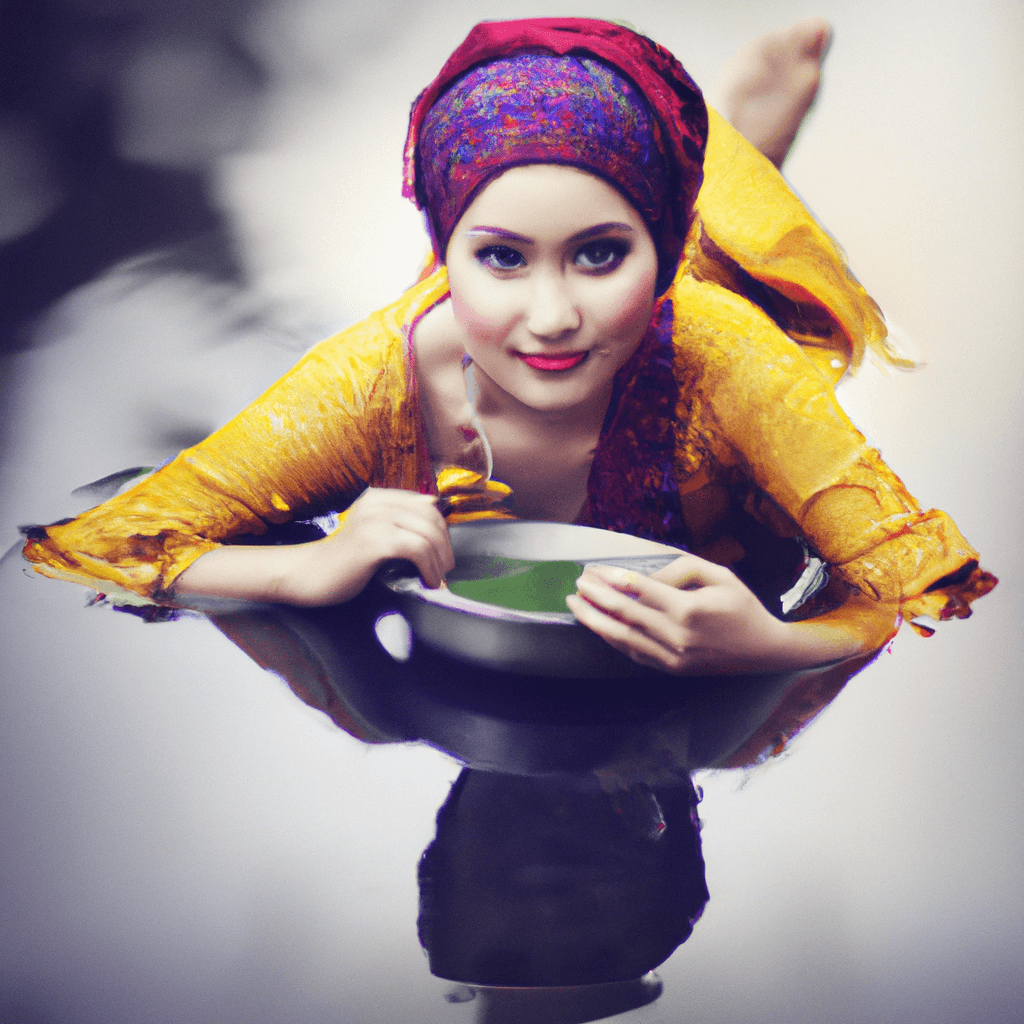}
  \caption{DALL·E prompt 5 (a modern Asian)}
  \label{girl_1}
\endminipage\hfill
\minipage{0.32\textwidth}
  \includegraphics[width=\linewidth]{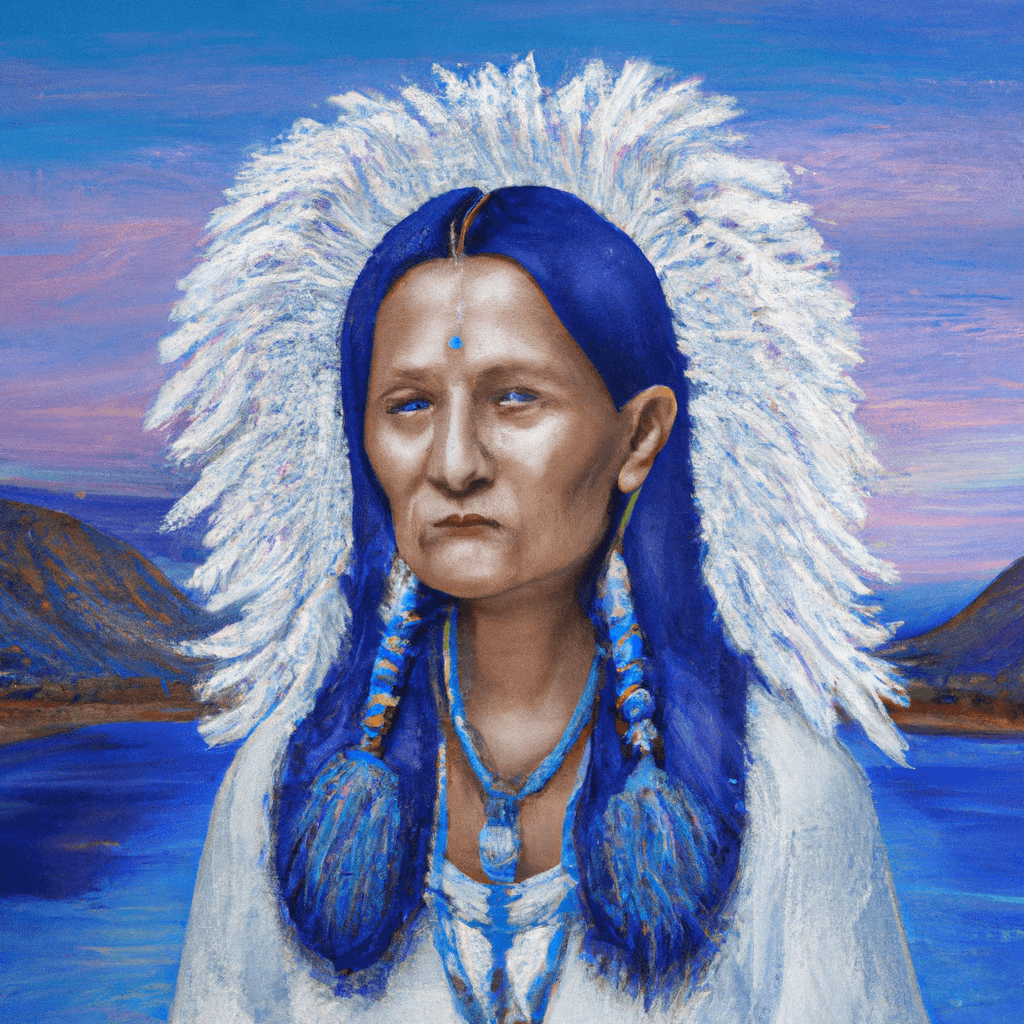}
  \caption{DALL·E prompt 5 (a native American)}
  \label{girl_2}
\endminipage\hfill
\minipage{0.32\textwidth}%
  \includegraphics[width=\linewidth]{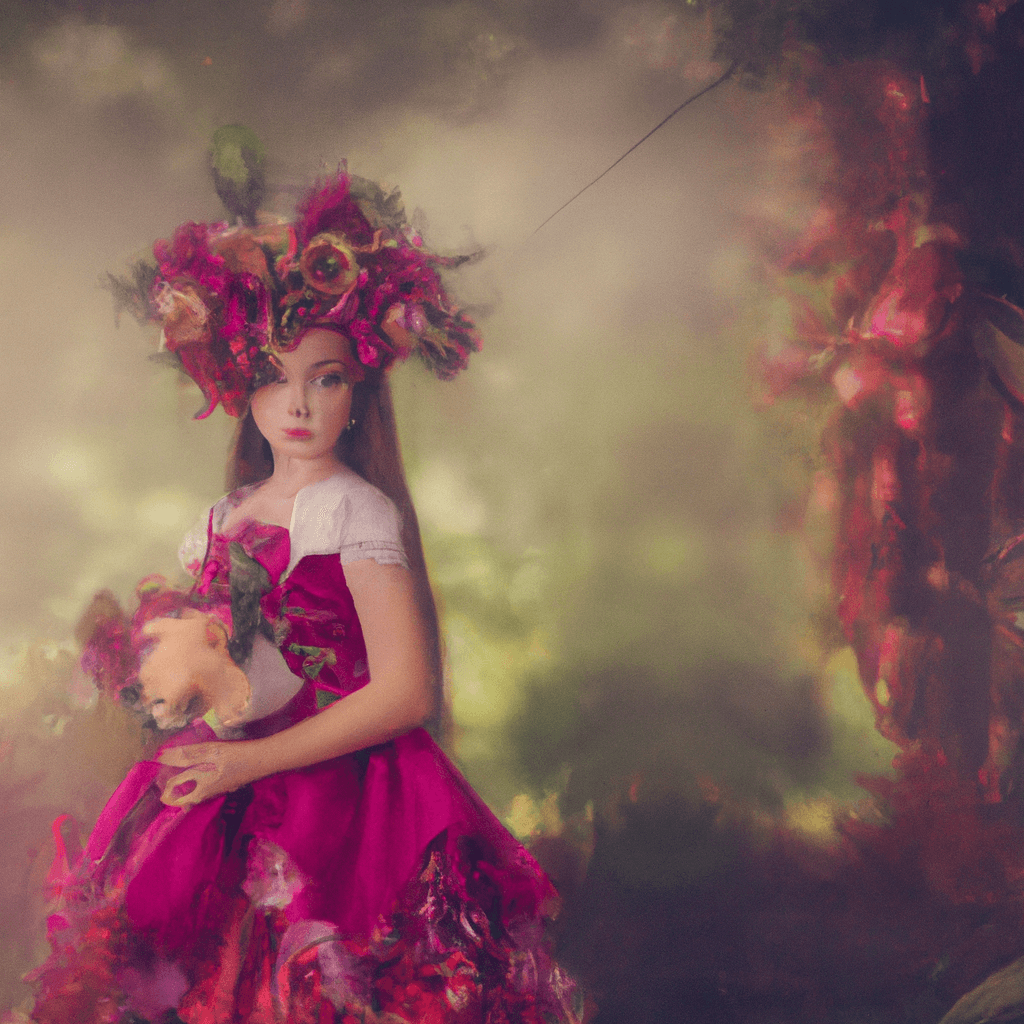}
  \caption{DALL·E prompt 5 (a lady in the Victorian era)}
  \label{girl_3}
\endminipage

\end{figure}

 Prompt 5 depicts mainly a photographic aesthetic image of a female character. This is evident in the prompt. However, the character's clothing and physical appearance do not match the biblical reference. Figure \ref{girl_1} could be from modern era while Figure \ref{girl_3} could be from Victorian era. There are also images of Native Americans (e.g. Figure \ref{girl_2}), More discussion is included in Section \ref{sec:discussion}.

Our research shows that DALL·E is the least accurate generator for producing these biblical prompts compared to the others. With a lack of accuracy but for some prompts, there are biblical references in the images even though the references are taken out of context of the prompt. 

\subsection{Midjourney}

Images generated by Midjourney follow a realistic style. Figures \ref{fig:last4} and \ref{fig:last5} are good examples to demonstrate its capability. Although there are flaws in the use of light, human characters and objects in the scene are presented with reasonable lighting settings. The generated images often have balanced composition with much attention paid to details.

\begin{figure}[!ht]
\minipage{0.32\textwidth}
  \includegraphics[width=\linewidth]{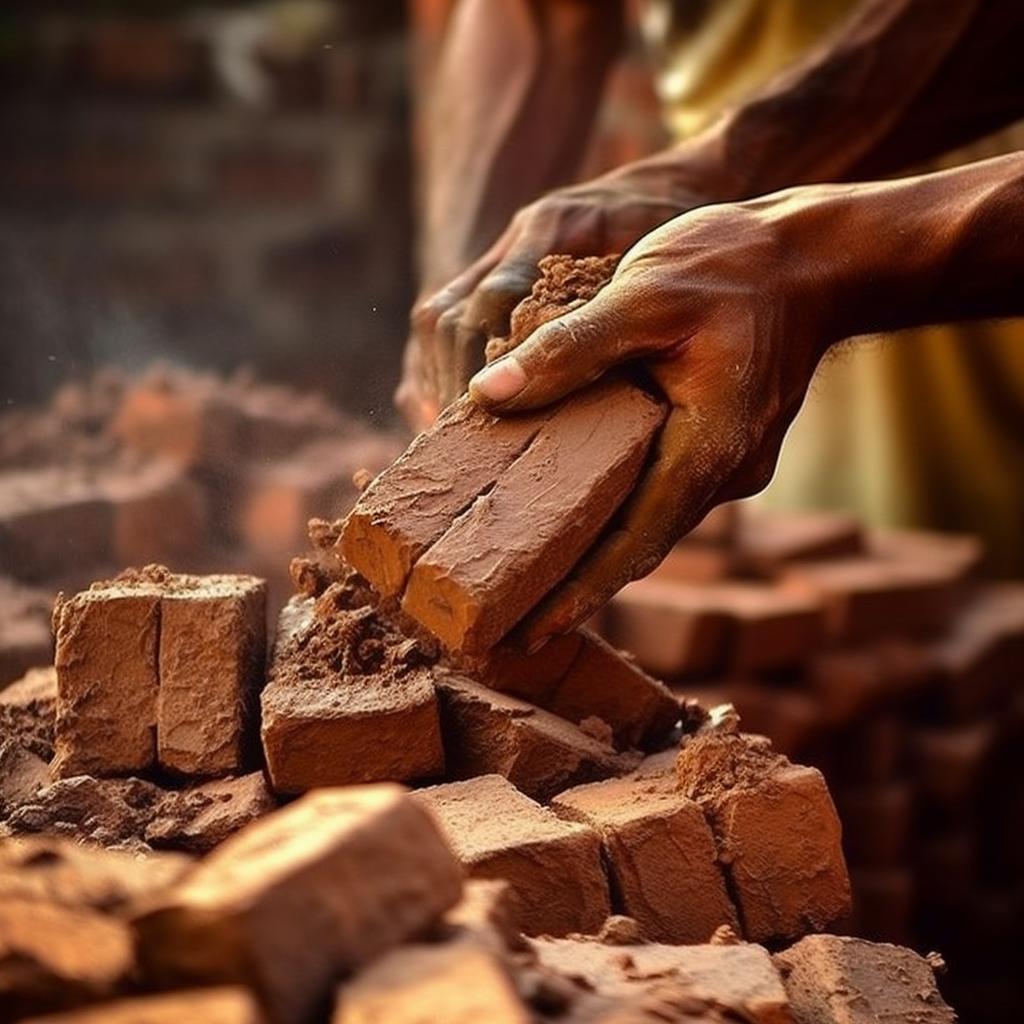}
  \caption{The details of hands in an image generated by Midjourney using prompt 2  (constructing the Babel Tower)}
  \label{hand1}
\endminipage\hfill
\minipage{0.32\textwidth}
  \includegraphics[width=\linewidth]{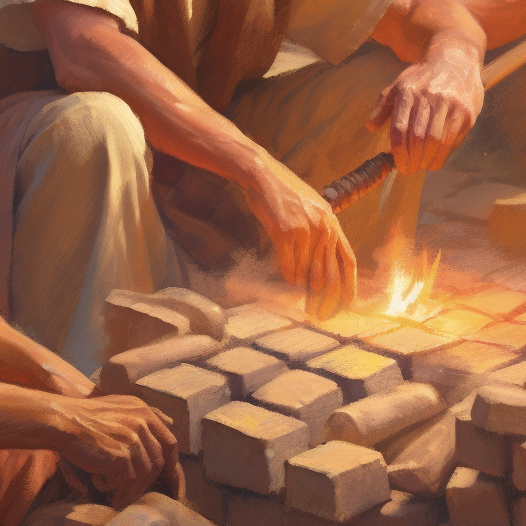}
  \caption{The details of hands in an image generated by Midjourney using prompt 2 (constructing the Babel Tower). The hand in the middle has overlapping structures}
  \label{hand2}
\endminipage\hfill
\minipage{0.32\textwidth}%
  \includegraphics[width=\linewidth]{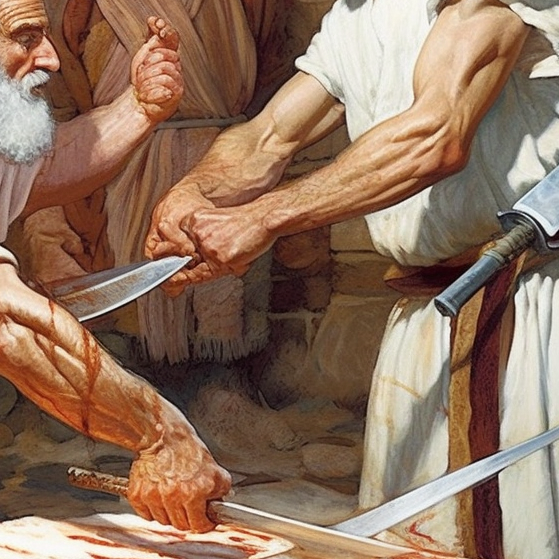}
  \caption{The details of hands in an image generated by Midjourney using prompt 3 (Abraham sacrifices his son). The hand on the bottom left has its wrist unnaturally bent. The structure of the three hands on the back are in disarray.}
  \label{hand3}
\endminipage

\end{figure}

Hands are generally considered among the hardest objects for art students to draw. Midjourney is fine-tuned towards generating realistic images, especially hands. It outperforms DALL·E on this task. Some examples are included in Figures \ref{hand1}, \ref{hand3}, and \ref{hand3}. However, it is still facing problems of often generating hands with more than five fingers or that seems to be the overlapping of multiple hands. This flaw does not affect the overall performance in generating the most realistic hands compared to the other generators.

\begin{figure}[!ht]
\minipage{0.32\textwidth}
  \includegraphics[width=\linewidth]{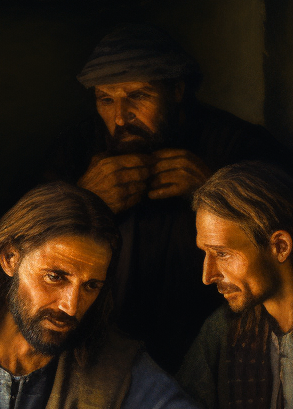}
  \caption{The facial details of human characters in a generated image by Mijourney using prompt 4 (the Last Supper)}\label{emotion 1}
\endminipage\hfill
\minipage{0.32\textwidth}
  \includegraphics[width=\linewidth]{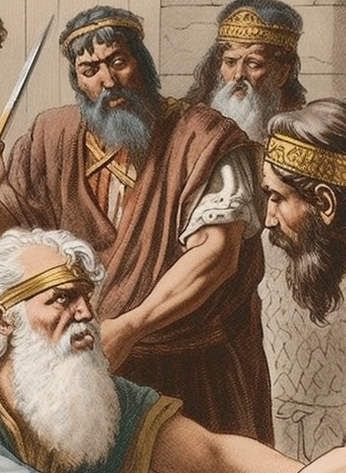}
  \caption{The facial details of human characters in a generated image by Mijourney using prompt 3 (Abraham sacrifices his son)}\label{emotion 2}
\endminipage\hfill
\minipage{0.32\textwidth}%
  \includegraphics[width=\linewidth]{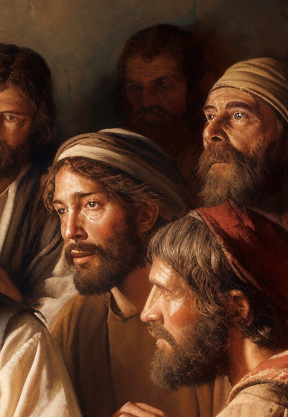}
  \caption{The facial details of human characters in a generated image by Mijourney using prompt 4 (the Last Supper)}\label{emotion 3}
\endminipage
\label{fig: faces}
\end{figure}

Midjourney also has consistency in generating facial details. This leads to characters having detailed emotions visible on their face. Figures \ref{emotion 1}, \ref{emotion 2} and \ref{emotion 3} show the detail in the faces of the characters depicted,  reflecting the intense mood.\footnote{These subtle facial details can be seen in the detail but can not be captured in the current workflow's sentimental model since we only look at pixels. Therefore, this aesthetic analysis carries some important weight in analyzing aspects of the image not captured by automated analysis.}.

\begin{figure}[h!]
\minipage{0.32\textwidth}
  \includegraphics[width=\linewidth]{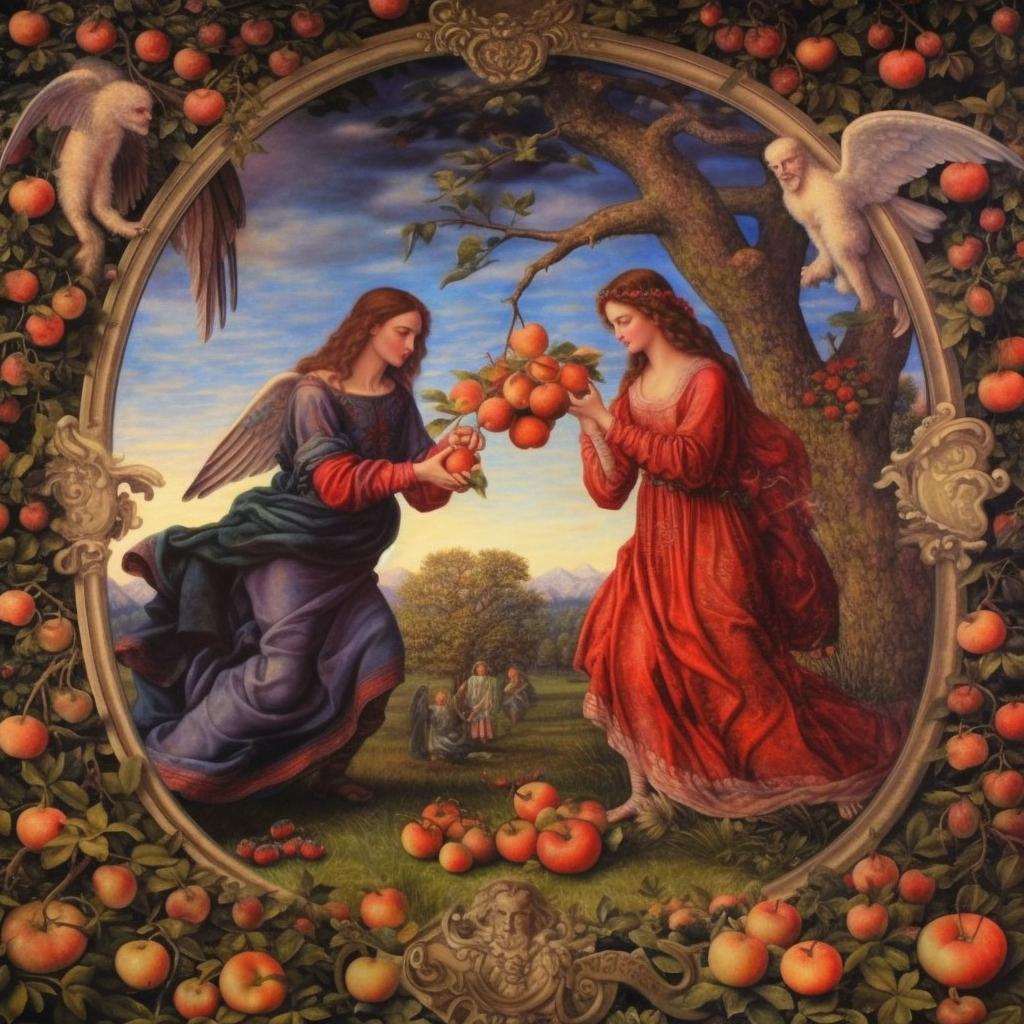}
  \caption{An image generated by Midjourney using prompt 1 (the expulsion from paradise) exhibits a similar pattern in composition as the other two with a garden arch, fruits (probably apples),  trees, and some creatures, which could be angles}\label{fig: pattern 1}
\endminipage\hfill
\minipage{0.32\textwidth}
  \includegraphics[width=\linewidth]{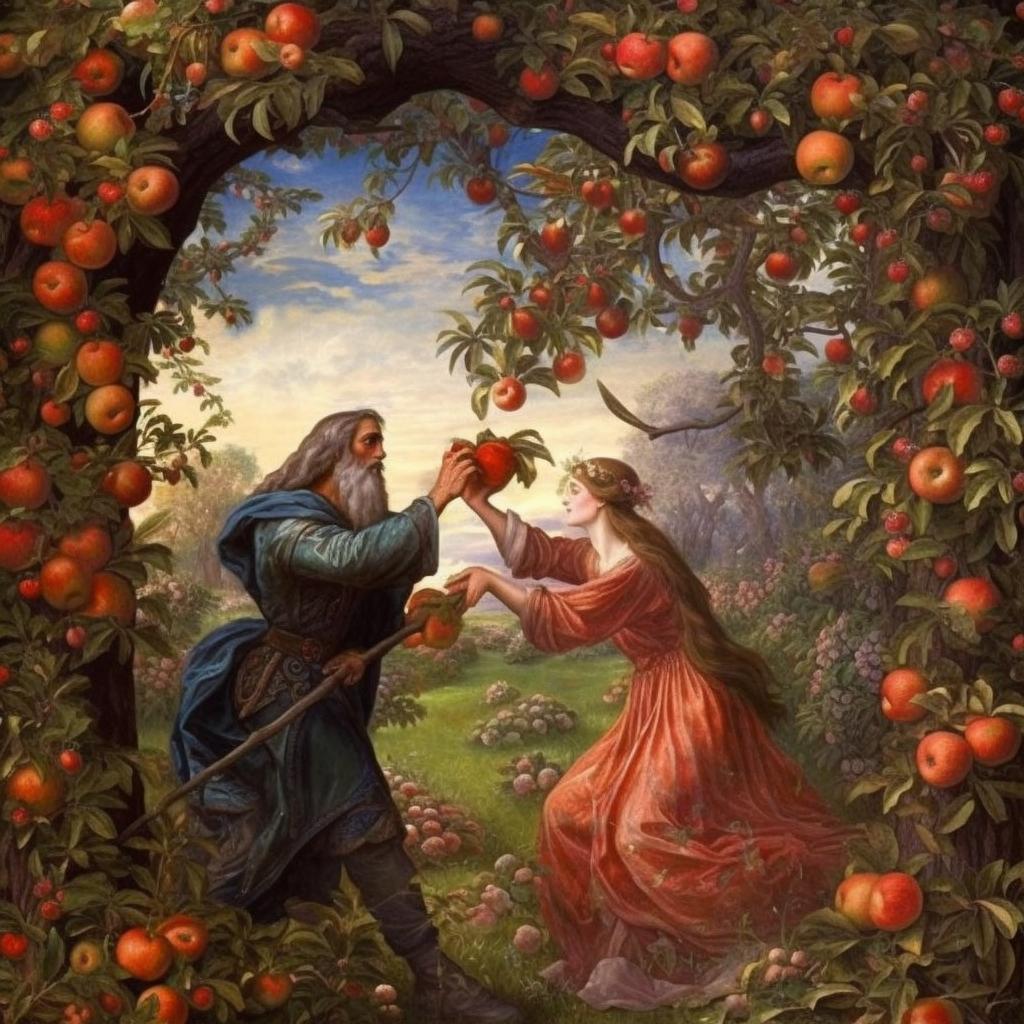}
  \caption{An image generated by Midjourney using prompt 1 (the expulsion from paradise) exhibits a similar pattern in composition as the other two with a garden arch, fruits (probably apples) and trees}\label{fig: pattern 2}
\endminipage\hfill
\minipage{0.32\textwidth}
  \includegraphics[width=\linewidth]{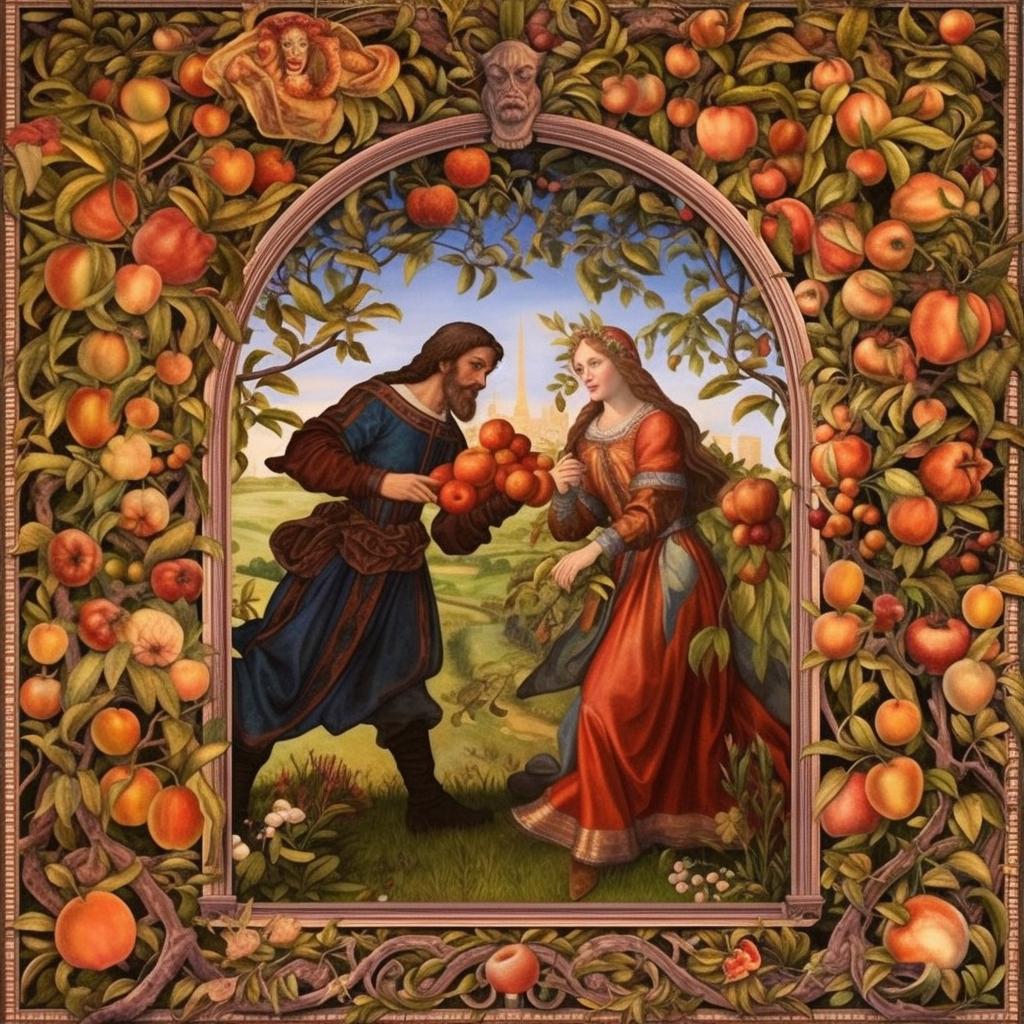}
  \caption{An image generated by Midjourney using prompt 1 (the expulsion from paradise) exhibits a similar pattern in composition as the other two with fruits (probably apple) and trees}\label{fig: pattern 3}
\endminipage
\end{figure}

One issue with Midjourney is the limited diversity of images it produces. The images tend to focus on characters and lack change in scenes, which was observed in both DALL·E and Stable Diffusion. As a result, the images often share similarities in terms of composition, environment, weather, objects, and art style, as shown in the accompanying Figures \ref{fig: pattern 1}, \ref{fig: pattern 2} and \ref{fig: pattern 3}.

Regarding religious accuracy in Midjourney, the content produced provides the most realistic religious perspective for the majority of the prompts' generated images. These images resemble storybook illustrations, with historical attire, environments, and characters from the prompts. However, prompt 1 has some shortcomings in this regard. The religious accuracy is not as strong, and the characters depicted seem to be from Western countries seen in Figure \ref{fig: pattern 1}, and the attire is not historically accurate, which shows the limitation and Western bias for prompt 1 image generation. 
Overall, Midjourney shows a high level of aesthetic details in human anatomy and objects in the images, but the trade-off comes from a lack of variety in the images. More discussion of the performance of Midjourney when compared with others is in Section \ref{sec:discussion}.

\subsection{Stable Diffusion}
\label{sec:StableDiffusion}
\begin{figure}[!ht]
\minipage{0.32\textwidth}
  \includegraphics[width=\linewidth]{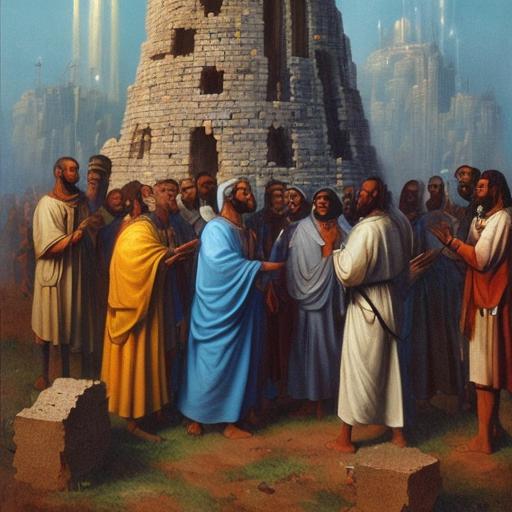}
  \caption{An image generated by Stable Diffusion using prompt 2 (constructing the Babel Tower). The image could be about some Africans with a background much like that in science fiction}\label{fig: var 1}
\endminipage\hfill
\minipage{0.32\textwidth}
  \includegraphics[width=\linewidth]{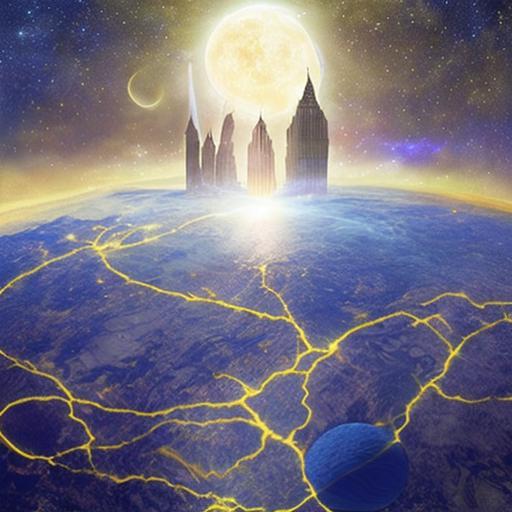}
  \caption{An image generated by Stable Diffusion using prompt 2 (constructing the Babel Tower). The image exhibits clear traits of science fiction.}\label{fig: var 2}
\endminipage\hfill
\minipage{0.32\textwidth}
  \includegraphics[width=\linewidth]{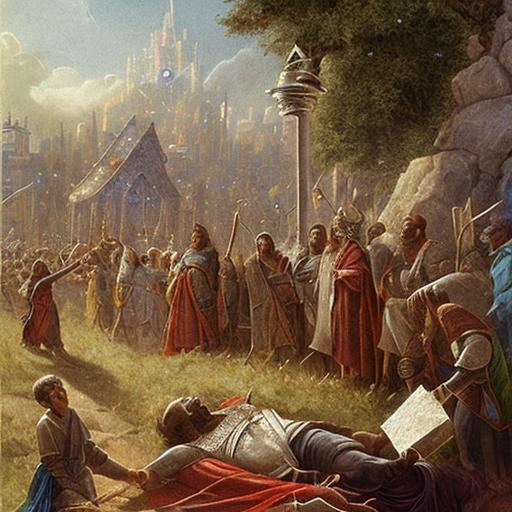}
  \caption{An image generated by Stable Diffusion using prompt 2 (constructing the Babel Tower). The image has a background much like that in science fiction}\label{fig: var 3}
\endminipage
\end{figure}

It appears that Stable Diffusion combines the styles of both DALL-E and Midjourney, resulting in unique images without specific themes. This allows for the generation of science fiction and fantasy themes. For instance, Prompt 2 showcases these themes as depicted in the accompanying Figures \ref{fig: var 1}, \ref{fig: var 2} and \ref{fig: var 3}.

\begin{figure}[h!]
\minipage{0.32\textwidth}
  \includegraphics[width=\linewidth]{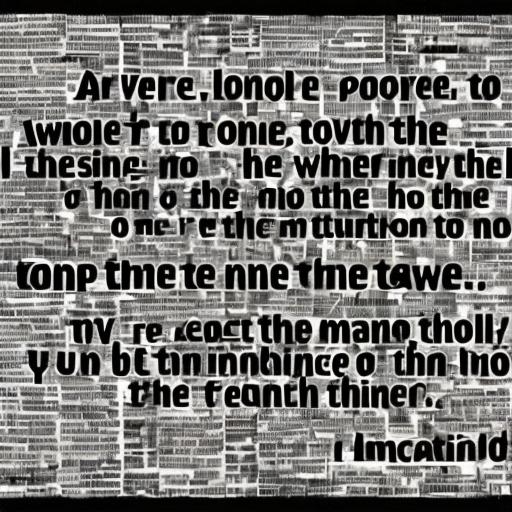}
  \caption{An image generated by Stable Diffusion (NS) using prompt 1 (the expulsion from paradise) exhibits features like that of Dall E}\label{fig:diverse1}
\endminipage\hfill
\minipage{0.32\textwidth}
  \includegraphics[width=\linewidth]{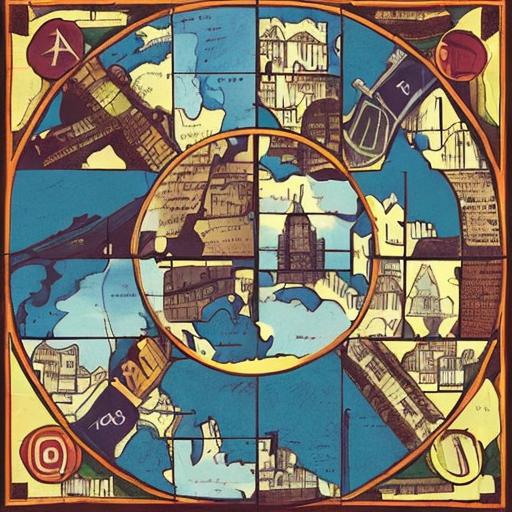}
  \caption{An image generated by Stable Diffusion (NS) using prompt 1 (the expulsion from paradise) in a different style}\label{fig:diverse2}
\endminipage\hfill
\minipage{0.32\textwidth}
  \includegraphics[width=\linewidth]{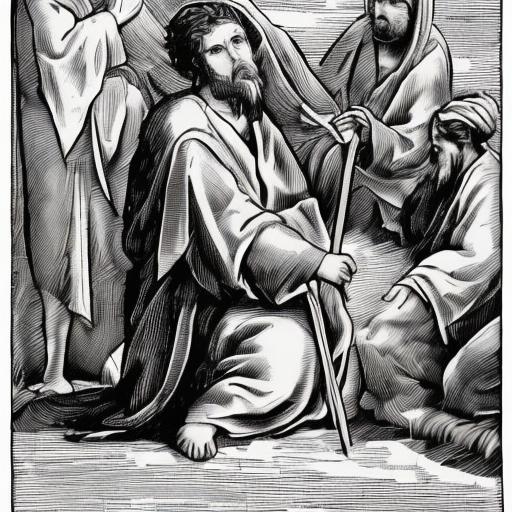}
  \caption{An image generated by Stable Diffusion (NS) using prompt 2 (constructing the Babel Tower) is in the style of a sketch }\label{fig:diverse3}
\endminipage
\end{figure}

The images produced exhibit a diverse range of aesthetic styles, lacking any fixed standard. Figures \ref{fig:diverse1}, \ref{fig:diverse2}, and \ref{fig:diverse3} are a few of such examples.  While some images accurately acknowledge religious prompts with halos and crosses, others fall short in their religious accuracy. See the discussion in Section \ref{sec:discussion} for examples.

\begin{figure}[h!]
\minipage{0.45\textwidth}
  \includegraphics[width=\linewidth]{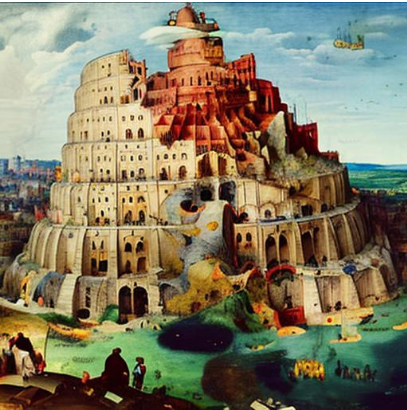}
  \caption{An image generated by Stable Diffusion using prompt 2 (construction of the Babel Tower)}\label{fig:tower_SD}
\endminipage\hfill
\minipage{0.53\textwidth}
\includegraphics[width=\linewidth]{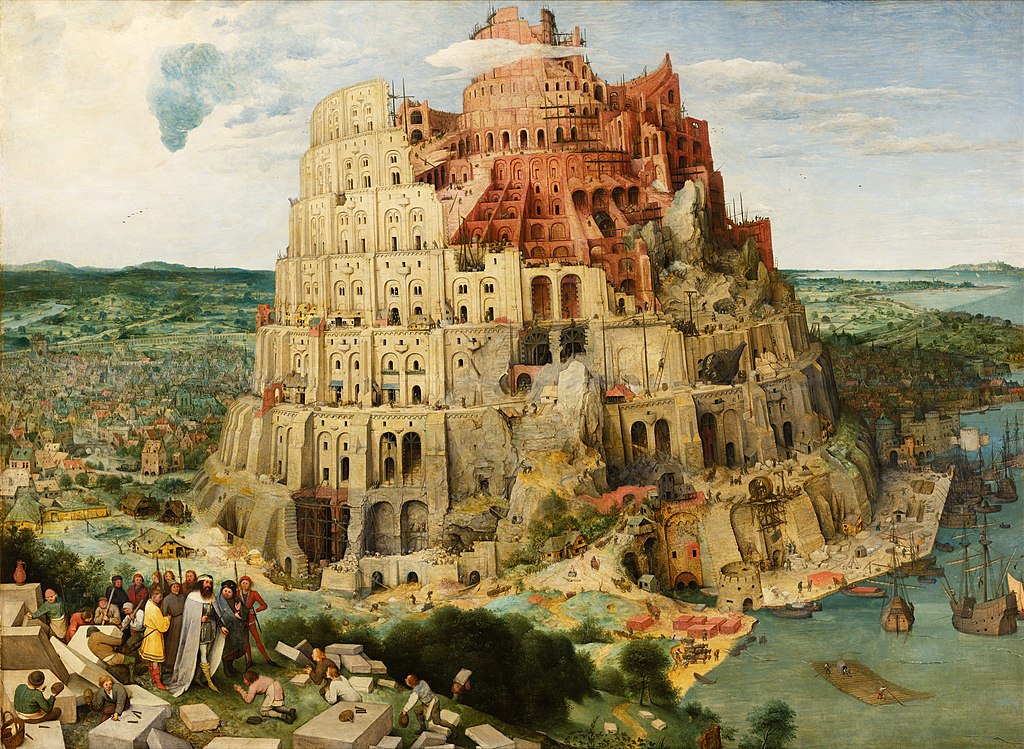}
   \caption{Pieter {\color{black}Bruegel}, Tower of Babel (1563)}\label{fig:tower_peter}
\endminipage\hfill
\end{figure}

One of the drawbacks of Stable Diffusion is that it can create difficulty in accurately depicting human autonomy, such as eyes and hands. Additionally, some of the resulting images are highly similar to the reference image, showing little difference. An example of this can be seen in Figures \ref{fig:tower_SD} and \ref{fig:tower_peter},  which shows that Bruegel's Tower of Babel painting was taken as a reference. 

Overall Stable Diffusion gives the most comprehensible versatile results for the prompt. It does not have an aesthetic, artistic style or level of detail as Midjourney, but it creates images that are more detailed than DALL·E and more versatile than Midjourney. However, it has less detail than Midjourney. The difference between its variants is to be discussed in the following sections.

\section{Automated Evaluation}
\label{sec:evaluation}

Next, we provide details on the evaluation and compare the scores in different settings as introduced in Section \ref{sec:methodology}. Given that the paper is aimed at digital humanities and theology researchers, the remaining text in this section outlines each simple metric rather than providing details with mathematical notions. We provide an example for each metric. More details of these metrics can be found in the supplementary material in the code repository. Since the evaluation of religious aspects and aesthetics cannot be performed completely automatically, they are manually assessed and included in Section \ref{sec:discussion}. 

\subsection{Measures}
\label{sec:measures}

Next, we introduce some measures for each assessed aspect. As a demonstration of feasibility, we propose a method for summarizing these measures for an overall evaluation. To do this, we first unify the outputs from the models used for each measure into a score in the interval [0, 1] with 0 being the minimum (e.g., nobody in an image) and 1 being the maximum (e.g,. the maximum number of people in all images). The results of each of the following measures that assess different aspects are then aggregated by computing the average across all the aspects assessed. Next, we provide details of the measures and explain with examples.\footnote{More details and intermediate results are provided in the supplementary material.} 

\textbf{Number of people} Recall that for each generated image, we obtain the number of people detected as described in Section \ref{sec:human-recognition}. We then take the mean for each generator. Moreover, we compute its ``distance'' from a selected painting using the absolute value of the difference in the number of people in it. We then compute the standard deviation of these distance values. This number represents the difference in the average number of people between the generated artwork and that of the selected human art. The lower these numbers are, the more similar these images are with respect to the number of humans. For example, among all the 50 generated images, the number of people detected is 5, 5, 6, $\ldots$, 10. The maximum number of people detected among all images is 16. We use N-Mean and N-STD to refer to the average number of people detected in the image and the measure of the standard deviation of differences with respect to the base. Then the unified average score would be $(\text{N-Mean}/50/16) = 5.4/16 = 0.34$. Assuming that the score for human artwork is 0.43 (a.k.a. the base reference), the final unified score would be the difference, which is 0.09. It is clear that the lower the score, the more similar the number of people is between the set of generated images and the selected human artwork. 

\textbf{Gender} Take the number of females for example: for each generated image, we compute the difference with the number of detected females regarding each painting. The average of the absolute value was taken and divided by the maximum number of females among all the generated images and paintings to unify this number to an interval of 0-1. The standard deviation can then be computed in a similar way as described above. For each generator, the resulting average and standard deviation for all its generated images for each prompt show the difference in the number of females as well as its diversity. That for males is computed similarly. We denote M-Mean and F-Mean for the mean of the number of females and males detected, respectively. F-STD and M-STD standards for the measure of the standard deviation of the difference concerning the paintings in the base, respectively. 

\textbf{Age}
 Since the detected ages are categorized into ten groups, we associate a number with each group (e.g. 1 for group 0-10, 2 for group 11-19). We then calculate the differences in each group between a given human artwork and a generated image. We then divide its absolute value by the maximum number of people detected. The subsequent procedure is similar to that of gender described above.

\textbf{Sentimental values and scores}
We compute two scores for sentimental analysis. For each generated image, we obtain the sentimental value as described in Section \ref{sec:sentimental}. Similar to the calculation steps as described above, {\color{black}  the sentimental value for each generator is the average of the difference in sentimental value between each generated image and painting for each prompt. For the second score, we compute for each patch\footnote{A patch is a section of $N \times N$ pixels in the image. In this work, for a selected painting and a generated image, we compute the difference in the sentimental value of patches at the same location. In this paper, we divide the images into $8 \times 8$ patches. The value is the average of such difference for all locations of the patches.} the sentimental value between the generated image and human art. The average of all the patches is then the score of this comparison. For each prompt, the overall \textit{sentimental score} for a generator is the average of the difference in the sentimental value of each pair of generated images and paintings.}

Finally, as a proof-of-concept, we take the overall score simply as the average of all the above-mentioned scores that measure the differences between generated images and human art. The lower the scores, the more the generated images are similar to the selected artwork.


\subsection{Number of humans and their gender}


Due to the page limit, we could not present the details of measures for each generator regarding each prompt. Instead, we present representative cases for evaluation. The overall results are summarized and discussed in Section \ref{sec:statistical-analysis-gender-age-number}.

\begin{figure}[!ht]
\centering
\includegraphics[width=\linewidth]{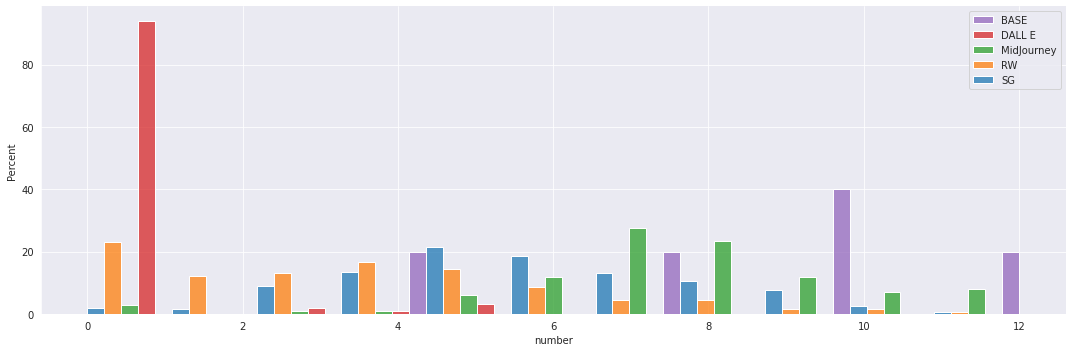}
\caption{The proportion (in percentage) of the number of people recognized in all generated images by selected generator for prompt 4 (the Last Supper). The red bar on the left shows that over 90\% of the image generated by DALL·E has no human character detected.}
\label{fig:distribution}
\end{figure}

 Figure \ref{fig:distribution} shows the proportion of recognized humans in the generated images for each generator regarding prompt 4 (the Last Supper). The green bars indicate that Midjourney can generate images with more humans while the images by DALL·E (shown in red) are very unlikely to have humans recognized. This is consistent with our findings in Section \ref{sec:method-DALLE}.  In this bar chart, we also present two variants of Stable Diffusion to show that the number of humans generated can vary between different versions. In comparison, more human characters can be found in human artwork. 
 
 This analysis shows that DALL·E lacks an understanding of the biblical context and does not recognize the characters in the prompt. In comparison, Midjourney and Stable Diffusion present some recognition of the biblical context, especially Midjourney.  Moreover, we noticed that, oftentimes, it is the case for Midjourney that a middle-aged man (representing Jesus) is around the center of the image with a few others surrounding him (e.g. Figure \ref{fig:last4} and Figure \ref{fig:last5}).

\begin{figure}[h]
 \centering
    \includegraphics[height = 8.5cm]{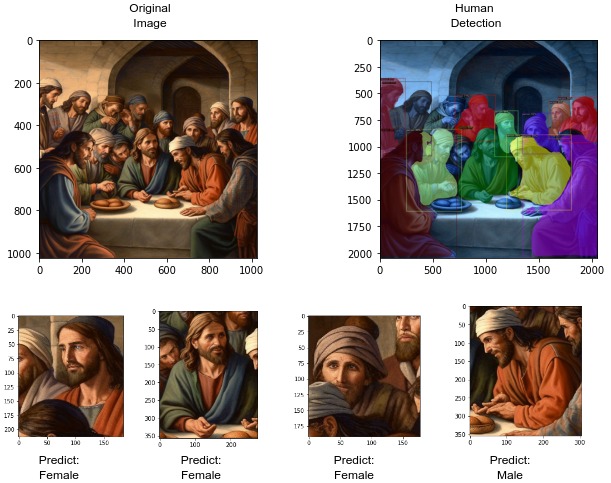}
    \caption{An example of human detection and gender classification using an image generated by Midjourney using prompt 4 (the Last Supper). 12 humans were recognized instead of 13. Three humans were misclassified as female.}
    \label{fig:example_limit_gender}
\end{figure}

Next, we reveal the limitation of our models by providing an example. In Figure \ref{fig:example_limit_gender}, the human character detection model we employed, Detectron2, was unable to identify all {\color{black}14} humans in the picture, only recognizing 12. This limitation may be due to the fact that most of the images used were {\color{black}not art pieces.} Detectron2 is primarily trained on photograph image types such as COCO, LVIS, and cityscapes, which could explain the reduced accuracy in artistic work. This limitation becomes more obvious as the realism of the art images decreases, making it more difficult for the model to identify human characters, {\color{black} especially if their faces are particularly hidden.} To increase the accuracy, one solution would be to use labelled artistic data to train the Detectron2 model in the future. {\color{black}We also noticed that the model we chose for gender detection has some limitations in its accuracy. Three humans, including the character in the middle, supposedly Jesus, was classified as female.}

\subsection{Age and gender}
{\color{black}Next, we study the accuracy by analyzing the age and gender of human characters in the generated images. In this subsection, we demonstrate by using prompt 1 (the expulsion of Adam and Eve from the paradise) and prompt 5 (the finding of Moses). We compare the results of Midjourney and DALL·E with three selected versions of Stable Diffusion. More detailed analysis and that of other prompts are included in the supplementary material.

\begin{figure}[!ht]
     \centering
     \begin{subfigure}[b]{0.3\textwidth}
         \centering
         \includegraphics[width=\textwidth]{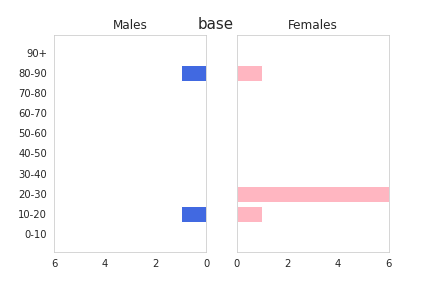}
         \caption{Base}
         \label{fig:base_pyramid_prompt0}
     \end{subfigure}
     \hfill
     \begin{subfigure}[b]{0.3\textwidth}
         \centering
         \includegraphics[width=\textwidth]{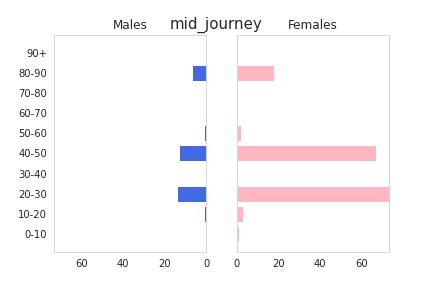}
         \caption{Midjourney}
         \label{fig:dall-E_pyramid_prompt0}
     \end{subfigure}
     \hfill
     \begin{subfigure}[b]{0.3\textwidth}
         \centering
         \includegraphics[width=\textwidth]{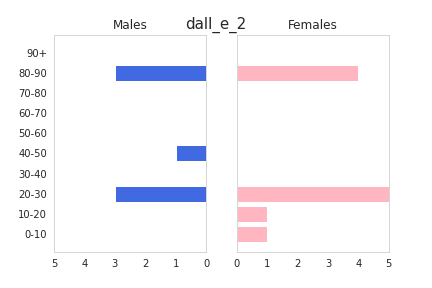}
         \caption{DALL·E 2}
         \label{fig:DALLE_pyramid_prompt0}
     \end{subfigure}
      \begin{subfigure}[b]{0.3\textwidth}
         \centering
         \includegraphics[width=\textwidth]{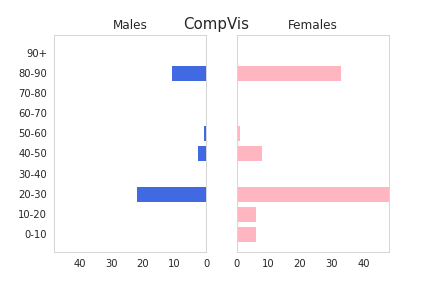}
         \caption{CV}
         \label{fig:CV_pyramid_prompt0}
     \end{subfigure}
     \hfill
     \begin{subfigure}[b]{0.3\textwidth}
         \centering
         \includegraphics[width=\textwidth]{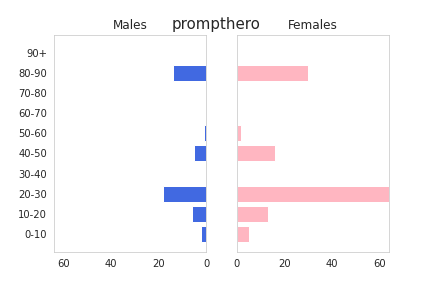}
         \caption{PH}
         \label{fig:PH_pyramid_prompt0}
     \end{subfigure}
     \hfill
     \begin{subfigure}[b]{0.3\textwidth}
         \centering
         \includegraphics[width=\textwidth]{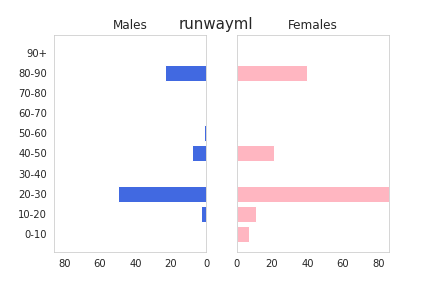}
         \caption{RW}
         \label{fig:RW_pyramid_prompt0}
     \end{subfigure}
        \caption{Population pyramids for prompt 1 (the expulsion of Adam and Eve from the paradise)}
        \label{fig:age-gender-prompt0}
\end{figure}

\begin{figure}[ht!]
\minipage{0.55\textwidth}
  \includegraphics[width=\linewidth]{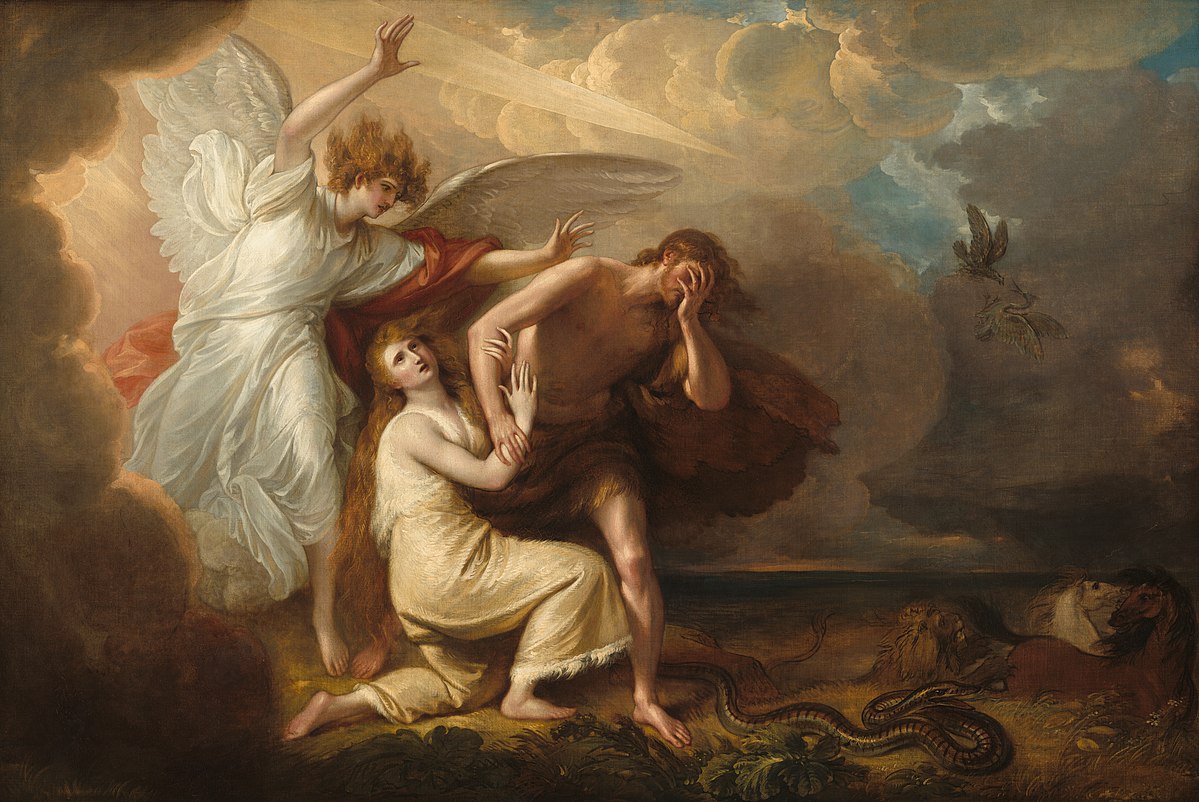}
  \caption{Cornelis van Poelenburg (1646)}\label{fig:adam-eve-face}
\endminipage\hfill
\minipage{0.40\textwidth}
\includegraphics[width=\linewidth]{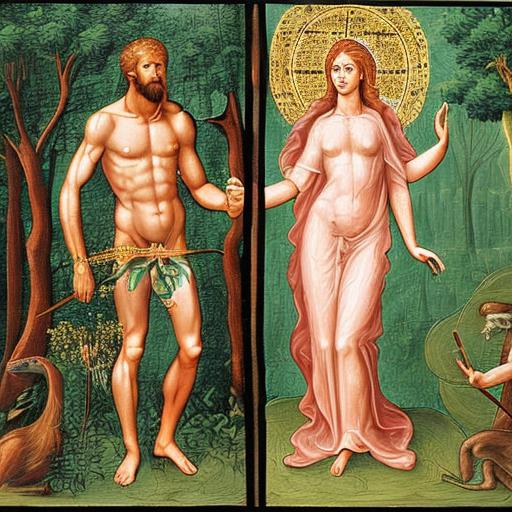}
   \caption{A generated image by PH using prompt 1 (the expulsion from paradise)}\label{fig:adam-eve}
\endminipage\hfill
\end{figure}

{\color{black}As for prompt 1, the population pyramids in Figure \ref{fig:age-gender-prompt0} show the distribution of males and females and their respective age categories (the counts are accumulated in all generated images and selected paintings). Adam and Eve are often presumed to be around the same age. It appears that the age groups are inclined to produce human characters in the group of 20-30 for all the generators, it can be observed that there is not an equal disparity of Male and Female characters detected.  However, the gender distribution appears to be predominantly female. A reason could be that in selected paintings, some male characters would hide their faces and weep (e.g. Figure \ref{fig:adam-eve-face}). This is almost never the case based on our manual examination of the generated images. Instead, male characters often do not show sorrow or sadness (see Figure \ref{fig: pattern 1},  \ref{fig: pattern 2},   \ref{fig: pattern 3}, and \ref{fig:adam-eve} for example). The limit of the model we used for gender detection was addressed in the subsection above. It was noticed that, when testing at a larger scale with other prompts, this issue arises as well. }

{\color{black} Next, we perform a similar analysis for prompt 5. The prompt includes some descriptions of female characters such as `girl',  `sister', `daughter', 
`mother', and `maidens'. In Figure \ref{fig:prompt5PopulationPyramid}, it was observed that a majority of female characters were generated, showing that the generators incorporate the semantics of character description in the text into the generated images.  
Moreover, we can also see that young women (20 to 30 years) are the main characters, which is also visible in the selected paintings.}

\begin{figure}[!ht]
     \centering
     \begin{subfigure}[b]{0.3\textwidth}
         \centering
         \includegraphics[width=\textwidth]{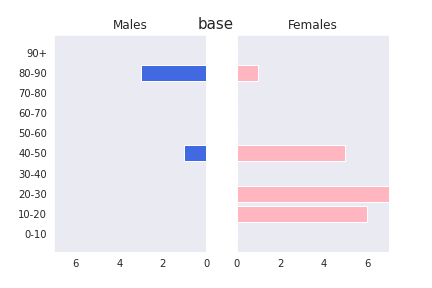}
         \caption{Base}
         \label{fig: }
     \end{subfigure}
     \hfill
     \begin{subfigure}[b]{0.3\textwidth}
         \centering
         \includegraphics[width=\textwidth]{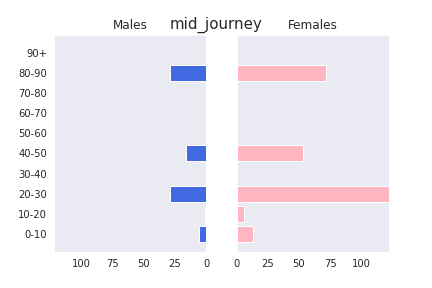}
         \caption{Midjourney}
         \label{fig:three sin x}
     \end{subfigure}
     \hfill
     \begin{subfigure}[b]{0.3\textwidth}
         \centering
         \includegraphics[width=\textwidth]{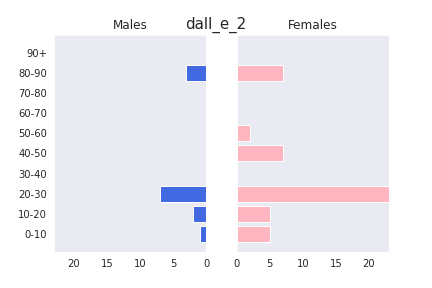}
         \caption{DALL·E 2}
         \label{fig:five over x}
     \end{subfigure}
      \begin{subfigure}[b]{0.3\textwidth}
         \centering
         \includegraphics[width=\textwidth]{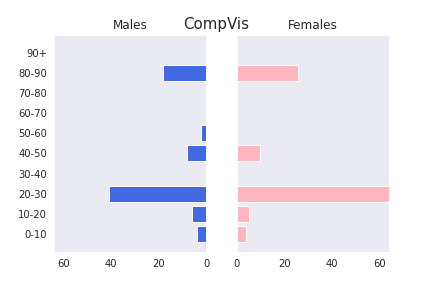}
         \caption{CV}
         \label{fig: }
     \end{subfigure}
     \hfill
     \begin{subfigure}[b]{0.3\textwidth}
         \centering
         \includegraphics[width=\textwidth]{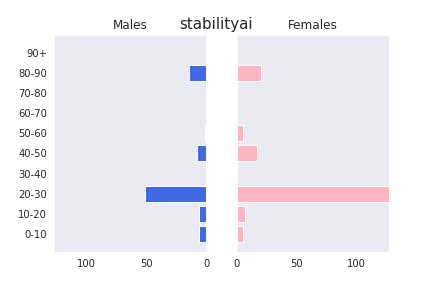}
         \caption{SAI}
         \label{fig:three sin x}
     \end{subfigure}
     \hfill
     \begin{subfigure}[b]{0.3\textwidth}
         \centering
         \includegraphics[width=\textwidth]{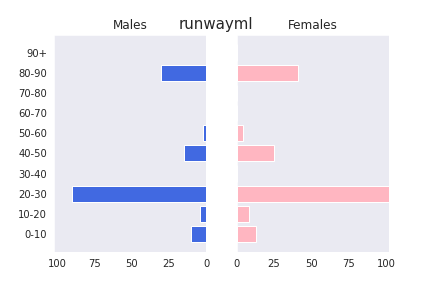}
         \caption{RW}
         \label{fig:five over x}
     \end{subfigure}
        \caption{Population Pyramids prompt 5 (the finding of Moses)}
        \label{fig:prompt5PopulationPyramid}
\end{figure}

\begin{figure}[!ht]
\minipage{0.32\textwidth}
  \includegraphics[width=\linewidth]{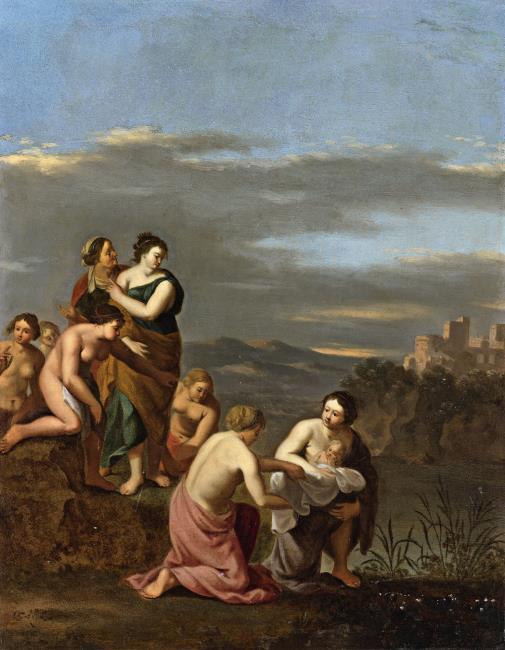}
  \caption{Toussaint Gelton  (1645 - 1680)}\label{fig:found1}
\endminipage\hfill
\minipage{0.32\textwidth}
\includegraphics[width=\linewidth]{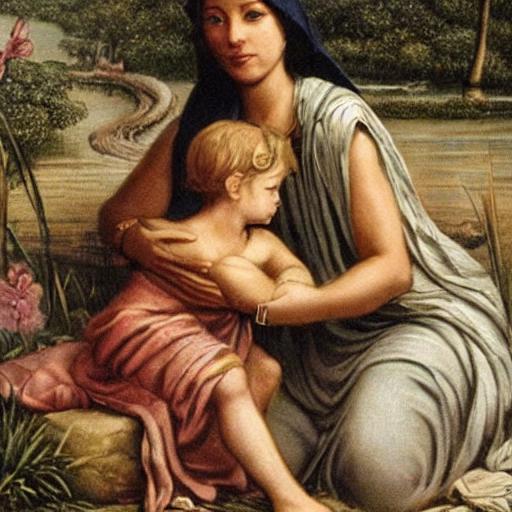}
   \caption{A generated image by RW using prompt 5 (Moses found)}\label{fig:found2}
\endminipage\hfill
\minipage{0.32\textwidth}
\includegraphics[width=\linewidth]{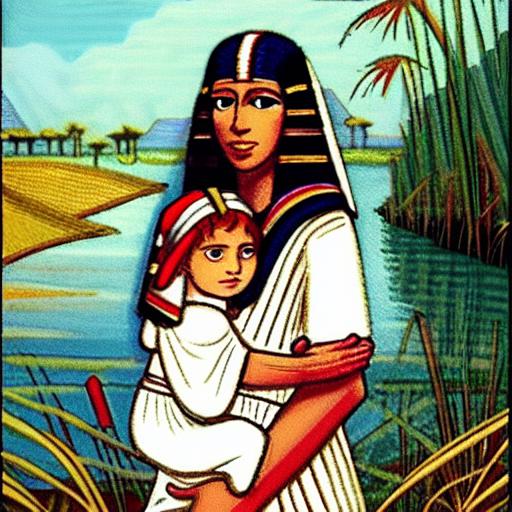}
   \caption{A generated image by DA using prompt 5 (Moses found)}\label{fig:found3}
\endminipage
\end{figure}

{\color{black}It was also noticed that no child (regardless of gender) was detected in our base. One of the reasons could be that the face of the baby is too small (see Figure \ref{fig:found1}). However, it was not the case for some generators. Figure \ref{fig:found2} and \ref{fig:found3} are two examples where the face of the child is clearly visable.}

\begin{table}[!ht]
\centering
\tiny
 \caption{Comparing the mean and standard deviation for the assessment of gender and number of people, as well as the overall accuracy for the generated images. M-Mean and F-Mean are the mean of the female and male detected, respectively. F-STD and M-STD standards for the measure of the standard deviation of the difference concerning the paintings in the base, respectively. N-Mean and N-STD refer to the average number of people detected in the image and the measure of the standard deviation of differences with respect to the base. For each row, the highest values are in bold font and the lowest values are underlined.}
\begin{tabular}{ll|l|l|l|l|l|l|l|l|l|l}
      &         & Base   & Midjourney     & DALL·E 2     & CV     & PH     & SAI             & DA     & NS     & SG              & RW              \\ \hline
\multicolumn{1}{l|}
{\multirow{6}{*}{Prompt 1}} & M-STD    & 0.5477 & 0.6364    & \multicolumn{1}{l|}{{\ul 0.1969}} & \multicolumn{1}{l|}{0.7423} & \multicolumn{1}{l|}{\textbf{0.8678}} & \multicolumn{1}{l|}{0.6781} & \multicolumn{1}{l|}{0.7635} & \multicolumn{1}{l|}{0.8257} & \multicolumn{1}{l|}{0.8149} & 0.7608 \\
\multicolumn{1}{l|}{}                          & M-Mean   & 1.4000   & 0.6212    & \multicolumn{1}{l|}{{\ul0.0400}} & \multicolumn{1}{l|}{0.8800} & \multicolumn{1}{l|}{1.1200} & \multicolumn{1}{l|}{0.6800} & \multicolumn{1}{l|}{\textbf{1.2300}} & \multicolumn{1}{l|}{0.9600} & \multicolumn{1}{l|}{1.0650} & 0.7100 \\
\multicolumn{1}{l|}{}                          & F-STD  & 0.8367 & \textbf{1.2500}    & \multicolumn{1}{l|}{{\ul 0.2611}} & \multicolumn{1}{l|}{0.6735} & \multicolumn{1}{l|}{0.7160} & \multicolumn{1}{l|}{0.8414} & \multicolumn{1}{l|}{0.8411} & \multicolumn{1}{l|}{0.6702} & \multicolumn{1}{l|}{0.8294} & 0.6574 \\
\multicolumn{1}{l|}{}                          & F-Mean & 0.8000   & 0.5682    & \multicolumn{1}{l|}{{\ul 0.0500}} & \multicolumn{1}{l|}{0.5300} & \multicolumn{1}{l|}{\textbf{0.9500}} & \multicolumn{1}{l|}{0.8250} & \multicolumn{1}{l|}{0.8600} & \multicolumn{1}{l|}{0.5550} & \multicolumn{1}{l|}{0.8400} & 0.4950 \\
\multicolumn{1}{l|}{}                          & N-STD  & 0.8367 & 0.8417    & \multicolumn{1}{l|}{{\ul 0.3786}} & \multicolumn{1}{l|}{0.8539} & \multicolumn{1}{l|}{0.8675} & \multicolumn{1}{l|}{0.9242} & \multicolumn{1}{l|}{0.9000} & \multicolumn{1}{l|}{0.9667} & \multicolumn{1}{l|}{0.9436} & \textbf{0.9889} \\
\multicolumn{1}{l|}{}                          & N-Mean & 2.2000   & 1.8712    & {\ul 0.0900}                      & 1.4100                      & 2.0700                      & 1.5050                      & \textbf{2.0900}                    & 1.5150                      & 1.9050                      & 1.2050 \\ \hline
\multicolumn{1}{l|}{\multirow{6}{*}{Prompt 2}} & M-STD    & 0.8944 & 1.9918    & \multicolumn{1}{l|}{{\ul 0.1407}} & \multicolumn{1}{l|}{1.5571} & \multicolumn{1}{l|}{1.4788} & \multicolumn{1}{l|}{\textbf{2.2337}} & \multicolumn{1}{l|}{0.4645} & \multicolumn{1}{l|}{0.9425} & \multicolumn{1}{l|}{1.4039} & 1.3423 \\
\multicolumn{1}{l|}{}                          & M-Mean   & 0.6000 & \textbf{2.5372}    & \multicolumn{1}{l|}{{\ul 0.0200}} & \multicolumn{1}{l|}{0.8600} & \multicolumn{1}{l|}{0.5700} & \multicolumn{1}{l|}{1.0250} & \multicolumn{1}{l|}{0.0800} & \multicolumn{1}{l|}{0.3100} & \multicolumn{1}{l|}{0.6700} & 0.4150 \\
\multicolumn{1}{l|}{}                          & F-STD  & 2.9155 & 0.9263    & \multicolumn{1}{l|}{{\ul 0.0000}} & \multicolumn{1}{l|}{1.2430} & \multicolumn{1}{l|}{1.2330} & \multicolumn{1}{l|}{1.7825} & \multicolumn{1}{l|}{0.1714} & \multicolumn{1}{l|}{1.0428} & \multicolumn{1}{l|}{\textbf{2.1416}} & 0.6111 \\
\multicolumn{1}{l|}{}                          & F-Mean & 2.0000 & 0.6033    & \multicolumn{1}{l|}{{\ul 0.0000}} & \multicolumn{1}{l|}{0.4800} & \multicolumn{1}{l|}{0.4300} & \multicolumn{1}{l|}{0.7800} & \multicolumn{1}{l|}{0.0300} & \multicolumn{1}{l|}{0.3050} & \multicolumn{1}{l|}{\textbf{0.9200}} & 0.2200 \\
\multicolumn{1}{l|}{}                          & N-STD  & 2.7018 & 2.3249    & \multicolumn{1}{l|}{{\ul 0.1407}} & \multicolumn{1}{l|}{2.5235} & \multicolumn{1}{l|}{2.3995} & \multicolumn{1}{l|}{\textbf{3.7183}} & \multicolumn{1}{l|}{0.5485} & \multicolumn{1}{l|}{1.7980} & \multicolumn{1}{l|}{3.2208} & 1.8050 \\
\multicolumn{1}{l|}{}                          & N-Mean & 2.6000 & {\textbf{3.1405}}    & \multicolumn{1}{l|}{0.0200} & \multicolumn{1}{l|}{1.3400} & \multicolumn{1}{l|}{1.0000} & \multicolumn{1}{l|}{1.8050} & \multicolumn{1}{l|}{{\ul 0.1100}} & \multicolumn{1}{l|}{0.6150} & \multicolumn{1}{l|}{1.5900} & 0.6350 \\ \hline
\multicolumn{1}{l|}{\multirow{6}{*}{Prompt 3}} & M-STD    & 0.8367 & 1.1114    & \multicolumn{1}{l|}{{\ul 0.4691}} & \multicolumn{1}{l|}{0.9574} & \multicolumn{1}{l|}{\textbf{1.5758}} & \multicolumn{1}{l|}{1.0206} & \multicolumn{1}{l|}{1.5062} & \multicolumn{1}{l|}{0.8876} & \multicolumn{1}{l|}{1.2703} & 0.9804 \\
\multicolumn{1}{l|}{}                          & M-Mean   & 0.8000 & \textbf{2.6956}    & \multicolumn{1}{l|}{{\ul 0.1100}} & \multicolumn{1}{l|}{1.0500} & \multicolumn{1}{l|}{2.0400} & \multicolumn{1}{l|}{1.4400} & \multicolumn{1}{l|}{2.2900} & \multicolumn{1}{l|}{0.8550} & \multicolumn{1}{l|}{2.1200} & 0.9400 \\
\multicolumn{1}{l|}{}                          & F-STD  & 0.4472 & 0.4842    & \multicolumn{1}{l|}{{\ul 0.3451}} & \multicolumn{1}{l|}{0.5807} & \multicolumn{1}{l|}{0.8689} & \multicolumn{1}{l|}{0.6841} & \multicolumn{1}{l|}{0.8771} & \multicolumn{1}{l|}{0.6386} & \multicolumn{1}{l|}{\textbf{0.8991}} & 0.5257 \\
\multicolumn{1}{l|}{}                          & F-Mean & 0.8000 & 0.2536    & \multicolumn{1}{l|}{{\ul 0.1100}} & \multicolumn{1}{l|}{0.3100} & \multicolumn{1}{l|}{0.5500} & \multicolumn{1}{l|}{0.3800} & \multicolumn{1}{l|}{\textbf{0.7200}} & \multicolumn{1}{l|}{0.3150} & \multicolumn{1}{l|}{0.5750} & 0.2450 \\
\multicolumn{1}{l|}{}                          & N-STD  & 1.1402 & 1.1160    & \multicolumn{1}{l|}{{\ul 0.6127}} & \multicolumn{1}{l|}{1.0398} & \multicolumn{1}{l|}{2.0156} & \multicolumn{1}{l|}{1.2750} & \multicolumn{1}{l|}{\textbf{1.6967}} & \multicolumn{1}{l|}{1.1033} & \multicolumn{1}{l|}{1.6015} & 1.1608 \\
\multicolumn{1}{l|}{}                          & N-Mean & 1.6000 & 2.9493    & \multicolumn{1}{l|}{{\ul 0.2200}} & \multicolumn{1}{l|}{1.3600} & \multicolumn{1}{l|}{2.5900} & \multicolumn{1}{l|}{1.8200} & \multicolumn{1}{l|}{\textbf{3.0100}} & \multicolumn{1}{l|}{1.1700} & \multicolumn{1}{l|}{2.6950} & 1.1850 \\ \hline
     
\multicolumn{1}{l|}{\multirow{6}{*}{Prompt 4}}   & M-STD   & 2.3021 & 2.1887          & {\ul 0.3258} & 2.0220 & 2.0220 & \textbf{2.2826} & 2.1082 & 1.5973 & 1.8842          & 2.0827          \\
\multicolumn{1}{l|}{}                                                                                                   & M-Mean  & 7.6000 & \textbf{6.9607} & {\ul 0.0700} & 2.1800 & 2.1800 & 3.3400          & 2.1400 & 1.4650 & 4.5850          & 2.6550          \\
\multicolumn{1}{l|}{}                                                                                                   & F-STD   & 0.7071 & 1.0151          & {\ul 0.2428} & 0.8954 & 0.8954 & \textbf{1.0677} & 0.6590 & 0.8896 & 0.9766          & 1.0296          \\
\multicolumn{1}{l|}{}                                                                                                   & F-Mean  & 2.0000 & \textbf{1.1372} & {\ul 0.0400} & 0.8100 & 0.8100 & 0.8400          & 0.5000 & 0.7500 & 1.0300          & 0.9850          \\
\multicolumn{1}{l|}{}                                                                                                   & N-STD  & 2.7928 & 2.3902          & {\ul 0.4691} & 2.4308 & 2.4308 & \textbf{2.7248} & 2.1997 & 1.8991 & 2.1887          & 2.5243          \\
\multicolumn{1}{l|}{}                                                                                                   & N-Mean & 9.6000 & \textbf{8.0980} & {\ul 0.1100} & 2.9900 & 2.9900 & 3.3400          & 2.6400 & 2.2150 & 5.6150          & 3.6400          \\ \hline
\multicolumn{1}{l|}{\multirow{6}{*}{Prompt 5}}                                                                          & M-STD   & 2.1213 & 0.8643          & {\ul 0.2777} & 0.7166 & 0.6887 & \textbf{0.7499} & 0.5222 & 0.3896 & 0.7421          & 0.6244          \\
\multicolumn{1}{l|}{}                                                                                                   & M-Mean  & 2.0000 & 0.3437          & {\ul 0.0600} & 0.5400 & 0.4800 & 0.5200          & 0.3000 & 0.1700 & \textbf{0.5450} & 0.4550          \\
\multicolumn{1}{l|}{}                                                                                                   & F-STD   & 0.8944 & 0.9636          & {\ul 0.5773} & 0.9519 & 0.9101 & 1.1072          & 0.7177 & 0.8084 & 0.8035          & \textbf{0.9664} \\
\multicolumn{1}{l|}{}                                                                                                   & F-Mean  & 1.4000 & \textbf{2.4765} & {\ul 0.5000} & 1.2700 & 2.0000 & 1.5100          & 1.5000 & 0.6400 & 1.7600          & 1.2750          \\
\multicolumn{1}{l|}{}                                                                                                   & N-STD  & 2.8809 & 0.9835          & 0.6407       & 1.1164 & 0.8466 & \textbf{1.2233} & 0.6030 & 0.8704 & {\ul 0.5599}    & 1.0783          \\
\multicolumn{1}{l|}{}                                                                                                   & N-Mean & 3.4000 & \textbf{2.8203} & {\ul 0.5600} & 1.8100 & 2.4800 & 2.0300          & 1.8000 & 0.8100 & 2.3050          & 1.7300          \\ \hline
\multicolumn{1}{l|}{\multirow{6}{*}{\begin{tabular}[c]{@{}l@{}}Overall \\ average\\ across all\\ prompts\end{tabular}}} & M-STD   & 1.3400 & 1.3585          & {\ul 0.2821} & 1.1991 & 1.3266 & \textbf{1.3930} & 1.0729 & 1.1701 & 1.2231          & 1.1581          \\
\multicolumn{1}{l|}{}                                                                                                   & M-Mean  & 2.4800 & \textbf{2.6317} & {\ul 0.0600} & 1.1020 & 1.2780 & 1.4010          & 1.2080 & 0.7520 & 1.7970          & 1.0350          \\
\multicolumn{1}{l|}{}                                                                                                   & F-STD   & 1.1602 & 0.9278          & {\ul 0.2853} & 0.8689 & 0.9247 & 1.0966          & 0.6533 & 0.8099 & \textbf{1.1300} & 0.6558          \\
\multicolumn{1}{l|}{}                                                                                                   & F-Mean  & 1.4000 & 1.0078          & {\ul 0.1400} & 0.6800 & 0.9480 & 0.8670           & 0.7220 & 0.5130 & \textbf{1.0250} & 0.6440          \\
\multicolumn{1}{l|}{}                                                                                                   & N-STD  & 2.0705 & 1.5313          & {\ul 0.4484} & 1.5929 & 0.8466 & \textbf{1.9731} & 1.1896 & 1.3275 & 1.7029          & 1.5115          \\
\multicolumn{1}{l|}{}                                                                                                   & N-Mean & 3.8800 & \textbf{3.7759} & {\ul 0.2000} & 1.7820 & 2.2260 & 2.1000          & 1.9300 & 1.2650 & 2.8220          & 1.6790         
\end{tabular}
\label{table:accuracy}
\end{table}

\subsection{Statistical analysis of number of people, gender, and age}
\label{sec:statistical-analysis-gender-age-number}

 Table \ref{table:accuracy} shows in detail the standard deviation (STD) and mean of the number of males and females, as well as the total number of people detected for prompts 4 and 5 in which males and females are dominant respectively so that the recognition of gender for Midjourney and Stable Diffusion has a similar mean to the human art. The standard deviation for male recognition is on average much higher across all generators for prompt 4. It is a recognized shortcoming in gender classification, as some long-haired male characters are classified as females (as illustrated in Figure \ref{fig:example_limit_gender}).}  
 {\color{black} Moreover, this analysis shows that the images generated using different versions of Stable Diffusion can differ significantly depending on the prompt. For example, for prompt 1, there are over two humans generated using DA on average, in contact with a much lower number of 1.2050 by RW. As for prompt 3, despite the fact that the average number of humans generated by PH and SG are similar, the significant difference in standard deviation shows that the diversity of images can be different. Overall, the last rows of Table \ref{table:accuracy} show that Midjourney has the highest number of people recognized. This number is close to that of our base of selected artworks. Some versions of Stable Diffusion have higher STDs, indicating that more diverse images were generated. DALL·E performs poorly in capturing human characters with its low score in both its standard deviation and mean values. More discussion of the accuracy is included in Section \ref{sec:discussion}.}

\subsection{Sentimental analysis} 
\label{sec:sentimental_analysis}

\begin{figure}[]
         \centering
    \includegraphics[scale= 0.35]{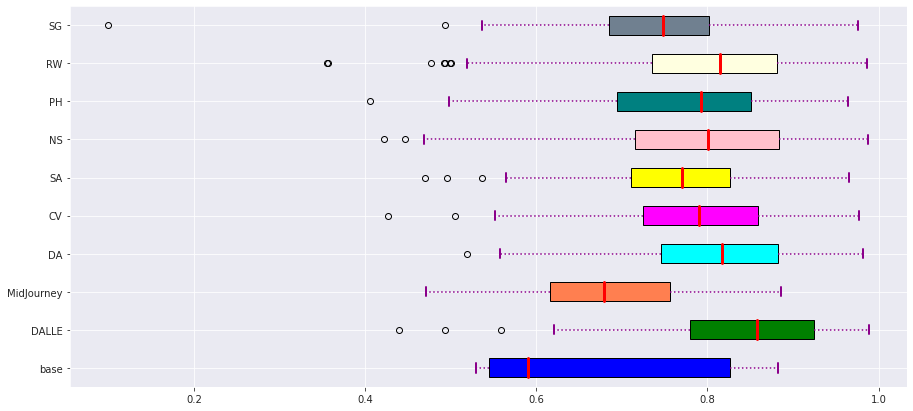}
     \caption{The sentimental value for prompt 2 by neural network models (0 represents negative and 1 positive)}
    \label{fig:senti}
\end{figure}

Next, we take prompt 2, for example, for sentimental analysis. As shown in Figure \ref{fig:senti}, the paintings seem to have a more neutral sentimental mean value than those of the generators. Compared with human art regarding the sentimental score (the difference between the sentimental value in human artwork and generated images), Midjourney captures the sentiment more accurately. Different versions of Stable Diffusion exhibit similar sentiments. In comparison, that of Dall·E has a higher overall value than the other generators. Table \ref{tab:sentimental_values_all} shows the sentimental score, which indicates that those by Midjourney are the most similar to the human artwork. In contrast, the sentimental score by DALL·E differs most from the paintings. 

\begin{table}[!ht]
\centering
\caption{\color{black}Comparing the average sentimental scores. For each row, the greatest values are in bold fonts and the lowest values are underlined.}
\begin{tabular}{l|l|l|l|l|l|l|l|l|l}
        & Midjourney & DALL·E 2 & CV       & PH       & SAI      & DA       & NS       & SG       & RW       \\ \hline 
Prompt 1 & {\ul 0.1042}  & \textbf{0.1660} & 0.1157 & 0.1295 & 0.1200 & 0.1293 & 0.1295 & 0.1574 & 0.1598 \\
Prompt 2 & {\ul 0.1522} & \textbf{0.1908} & 0.1777 & 0.1718 & 0.1658  & 0.1751 & 0.1834 & 0.1645 & 0.1897 \\
Prompt 3 & {\ul 0.1671} & 0.1816 & 0.1743 & 0.1724 & 0.1720 & \textbf{0.1902} & 0.1745 & 0.1767 & 0.1690 \\
Prompt 4 & 0.1491 & \textbf{0.1739} & 0.1588 & 0.1524 & 0.1512 & 0.1628 & 0.1687 & {\ul 0.1496} & 0.1516  \\
Prompt 5 & 0.1424& \textbf{0.1857} & 0.1474 & 0.140 & {\ul 0.1396} & 0.1651 &0.1754  &  0.1648 & 0.1708 \\
\end{tabular}
\label{tab:sentimental_values_all}
\end{table}



\subsubsection{Overall Scores}

The overall results per prompt can be seen in Table \ref{tab:prompt_table}. The smaller the number of the score, the more similar the generator is to the human paintings. Such higher similarity to human art performance is considered here as a better result (cf. Section \ref{sec:methodology-religious-aesthetics}). The score shows the average of all the calculated scores, giving us an indication of the overall difference between the human artwork against their respective AI counterparts.\footnote{For details analysis see supplementary analysis.} It shows that Midjourney is the most similar to the selected human artwork. Midjourney scores the best in the sentimental score for all prompts. In addition, the score for age, gender and number of people is also one of the best. {\color{black} The output of DALL·E differs the most from selected human art for prompt 1, 2, 4, and 5.} With it lacking the capabilities to produce recognizable human characters that the CNN model is able to detect and in turn not being able to score human characteristics. On the contrary, Stable Diffusion similarity depends on the prompt with some prompts producing more similar paintings to the human artwork. Its variations show a slight difference in score per prompt input. The last row in Table \ref{tab:prompt_table} is for the overall scores. It shows the average score tallied up from all prompts. The smaller the score, the better the performance. As we can see, Midjourney gives the best overall score. Its performance is significantly better than that of various versions of Stable Diffusion and outperforms DALL·E.

\begin{table}[]
\centering
\caption{Comparing the overall score of different generators for each prompt}
\begin{tabular}{l|l|l|l|l|l|l|l|l|l}
         & Midjourney & DALL·E 2 & CV       & PH       & SAI      & DA       & NS       & SG       & RW       \\ \cline{1-10} 
Prompt 1 & {\ul 0.1148}    & \textbf{0.1852} & 0.1208 & 0.1237 & 0.1261 & 0.1275 & 0.1285& 0.1417 & 0.1508 \\
Prompt 2 & {\ul 0.1196}    & \textbf{0.1384} & 0.1340 & 0.1309 & 0.1347 & 0.1302 & 0.1358 & 0.1322 & \textbf{0.1384} \\
Prompt 3 & 0.1219    & 0.1333 & 0.1219 & 0.1407 & 0.1223 & \textbf{0.1515} & 0.1245 & 0.1364 & {\ul 0.1216} \\
Prompt 4 & {\ul 0.1286}    & \textbf{0.2490} & 0.1926 & 0.1662 & 0.1743 & 0.2019 & 0.2087 & 0.1508 & 0.1788 \\
Prompt 5 & {\ul 0.1448} & \textbf{0.1883} & 0.1690 & 0.1613 & 0.1657 & 0.1698 & 0.1840 & 0.1597 & 0.1665 \\ \hline
Overall & {\ul 0.1279} & \textbf{0.1788} & 0.1477 & 0.1446 & 0.1446 & 0.1562 & 0.1563 & 0.1460 & 0.1512

\end{tabular}
\label{tab:prompt_table}
\end{table}


\section{Discussion}
\label{sec:discussion}

Our evaluation shows that Midjourney can generate illustration-like images that are most similar with respect to the features under evaluation when compared to our base of selected paintings. There could be several reasons for this. Images generated by Midjourney have more sophisticated details, which improves the accuracy of the recognition of objects. The details of faces, especially those partially hidden, covered, or under the shadow, can improve the detection of faces, which improves the detection of human characters in the generated images. This is particularly the case where multiple people are present, especially those using prompt 4. 
Some images by Midjourney capture the religious context with good composition and details of facial expressions, and thus could be  adopted with little modification as illustrations for biblical blogs, illustrated Bibles, etc. This corresponds to the interests of realistic details in the Renaissance and Baroque periods (cf. Sections \ref{sec:methodology-religious-aesthetics} and \ref{sec:religious-aesthetics}). An illustration of this are the anatomical details we find in both Renaissance paintings and Midjourney images. This may also be due to the inclusion of Renaissance art in its training data.  Stable Diffusion is different in that some images exhibit some level of abstraction. Some images could serve as inspiration for artists for further development. Although some generated images fail to capture the religious context and historical background, they could be useful for science fiction (e.g. Figure \ref{fig: var 2}). The DALL·E 2 images can vary significantly in style, which could reveal some degrees of creativity. This touches upon the general questions as to how we should define creativity in relation to accuracy. Do we assess the Tower of Babel depicted as a skyscraper or skyscrapers in the background of the Last Supper scene (Figure \ref{fig:sky}) as ``inaccurate'' or as ``creative'', or both? If creativity is defined in terms of something that is unique and unpredictable, it is not always easy to distinguish between ``inaccuracy'' and ``creativity''.\footnote{For the notion of digital creativity, see \cite{Peursen:2010}; for theological and anthropological consideration in relation of artificial creativity, see \cite{Klooster:2021}.}

One of the limitations is the number of humans detected in the images. In our experimental setting, despite that we took a relatively low confidence score of 0.8, under this setting, some human faces were not detected due to their imperfect quality, missing details, shadows, or partial coverage. This threshold could be lower in future work. 

Another the limitations of the model is its tendency to mistakenly recognize female characters even when they are obviously male. This is particularly the case for prompts 1, 3, and 4. Some physical attributes of long hair, robes, and head scarves, commonly seen in biblical times, are generalized by the model to be associated with the female gender. This may be due to ImageNet's training data, which mostly associates the above-mentioned features with women. This limitation affects the accuracy of gender classification in biblical characters, leading to misclassification in the detector despite their accurate representation in the images.

\begin{figure}[!ht]
\centering
\begin{subfigure}{.45\textwidth}
  \centering
  \includegraphics[width=.9\linewidth]{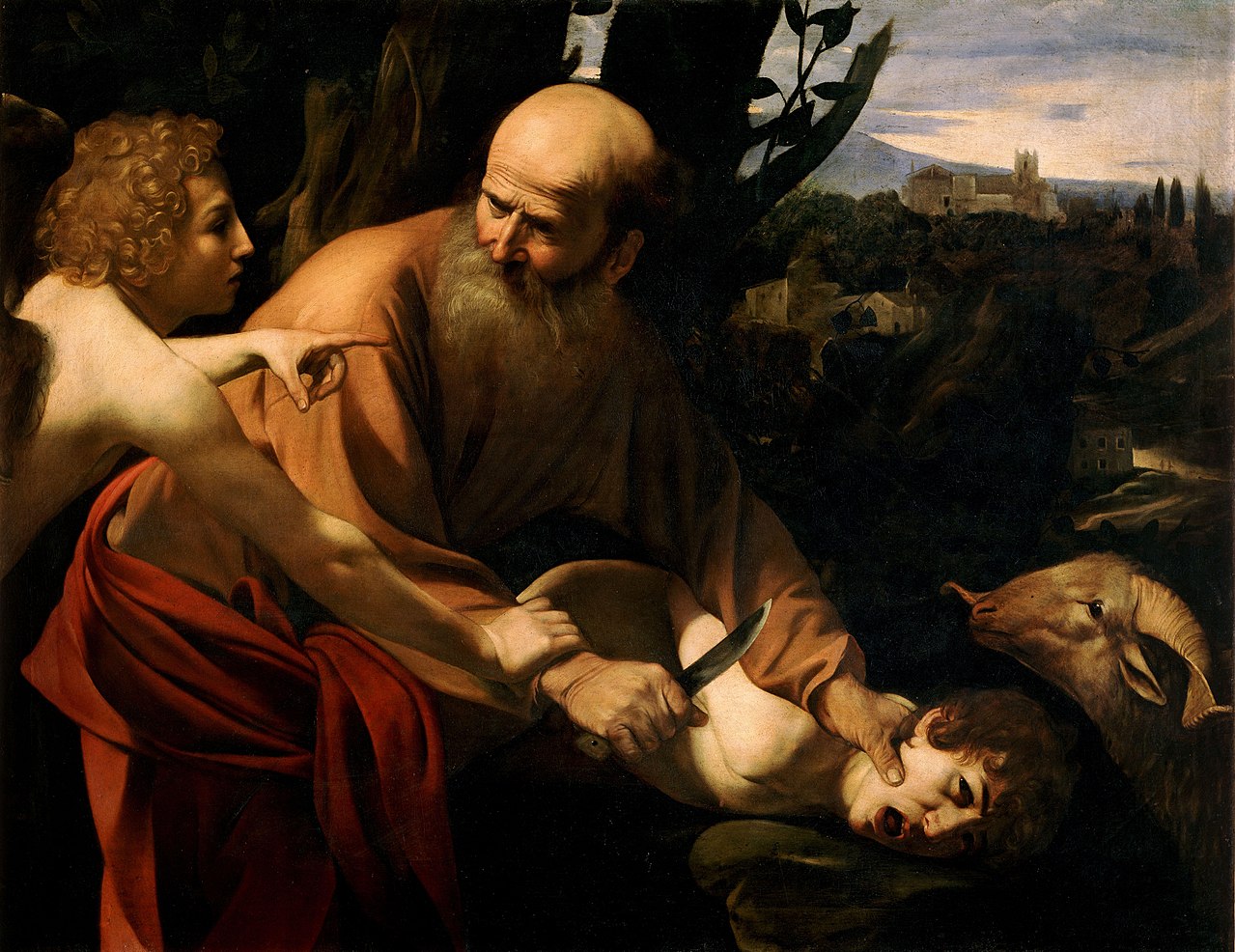}
  \caption{The Sacrifice of Isaac by Carvaggio (1603)}
  \label{fig:sub1}
\end{subfigure}%
\begin{subfigure}{.45\textwidth}
  \centering
  \includegraphics[width=.9\linewidth]{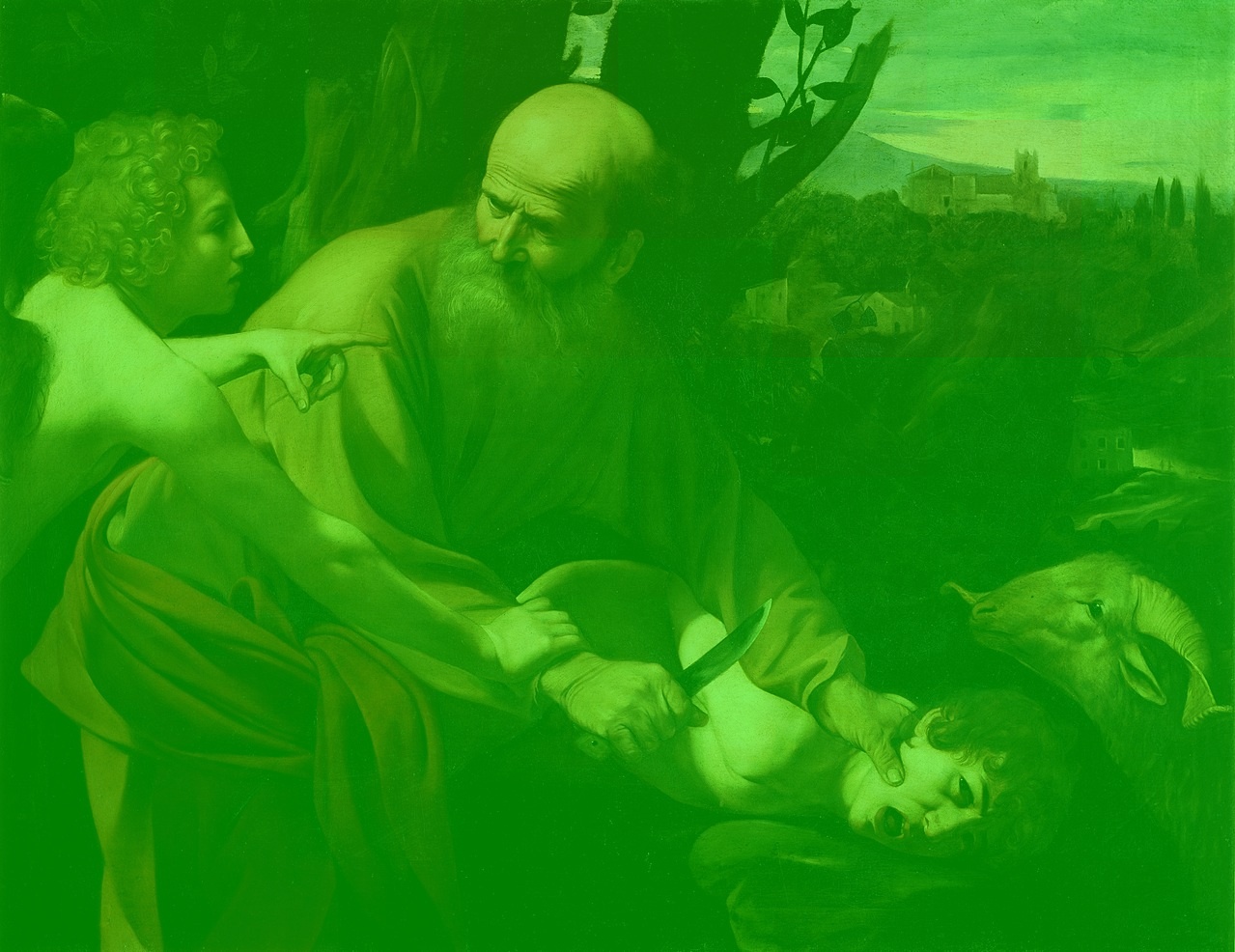}
  \caption{The corresponding heat map of sentiment values}
  \label{fig:sub2}
\end{subfigure}
\caption{An example how bright spots can change the sentimental value of a painting from negative to neutral}
\label{fig: abstract example}
\end{figure}

The sentimental model is not free from its limitations. Although it looks at the pixel brightness as a factor to depict the sentiment of the image, images that have a negative sentiment towards people, such as in Figure \ref{fig: abstract example} can be scored neutral. The score reflects the weakness of bright spots in the images, misleading the model to give a positive or neutral final score. Therefore, for a more accurate sentiment score, other aspects of the image have to be considered other than pixel brightness when performing sentimental evaluation.

\begin{figure}[ht!]
\minipage{0.32\textwidth}
  \includegraphics[width=\linewidth]{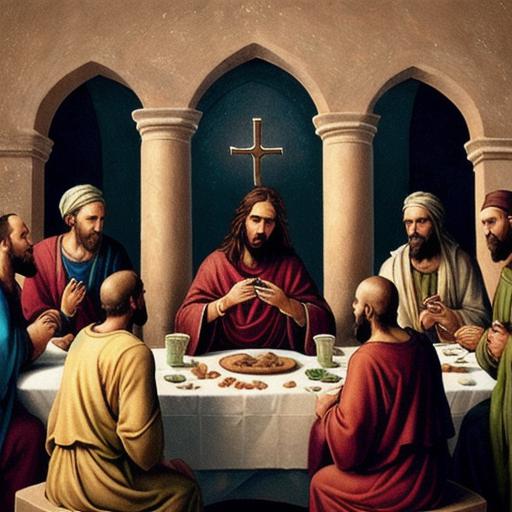}
  \caption{An image generated by Stable Diffusion (SG) with a cross on the wall for prompt 4 (the last supper)}\label{fig:cross1}
\endminipage\hfill
\minipage{0.32\textwidth}
  \includegraphics[width=\linewidth]{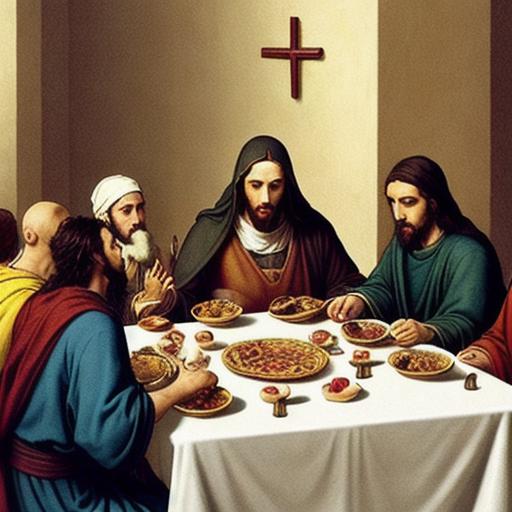}
  \caption{An image generated by Stable Diffusion (SG) with a cross on the wall for prompt 4 (the last supper)}\label{fig:cross2}
\endminipage\hfill
\minipage{0.32\textwidth}
  \includegraphics[width=\linewidth]{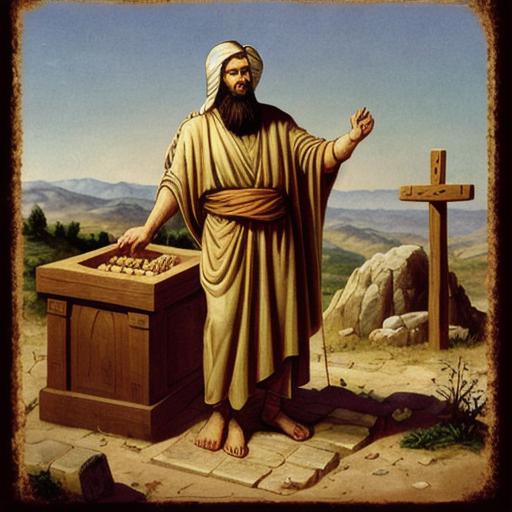}
  \caption{An image generated by Stable Diffusion (SG) using prompt 3 (Abraham sacrifices his son, Old Testament)}\label{fig:cross3}
\endminipage
\end{figure}

More details do not always imply better accuracy. We noticed, for example, that \textit{the cross} can be shown in the same generated image as Jesus. As a symbol of the Christian faith, this is understandable, but displaying a cross in a scene preceding Jesus' death is anachronistic. For example, Figures \ref{fig:cross1} and \ref{fig:cross2} include the cross for scenes about the last supper. Figure \ref{fig:cross3} presents the cross in images corresponding to prompt 3, which corresponds to the Old Testament.

Details can also lead to challenges at the detail level. Despite that the version of Midjourney has been fine-tuned for the generation of hands, none of the generators can generate perfect hands. We often observe polydactyly (one or multiple supernumerary fingers). For all generators, imperfections were observed on objects such as wings, plates, etc. The analysis of the style of glasses (some are very modern), interior design, and clothes is another issue that requires expertise beyond the authors' capacity.

 In this work, we do not implement automated analysis of text in our workflow. Thus, the semantic correspondences were analyzed manually but could be done with automation in the future (e.g. when scaling up to thousands of prompts). We noticed that the differences are significant: DALL·E seems to be unable to make sense of some text prompts and their context. Midjourney and Stable Diffusion perform better, but differently. In the case of the Tower of Babel (prompt 2), the training of Stable Diffusion seems to rely on traditional paintings of the biblical scene, whereas Midjourney picks up the building activity in a relatively naturalistic way (like cartoons or illustrations in children’s Bibles).
 
 Some semantic aspects of the text prompts apparently posed challenges. In the case of the Last Supper (prompt 4), the designation "the twelve" referring to the twelve disciples is in most cases not picked up. The text prompt contains some concrete objects (cup, bread), but also much conversation. This may have evoked the confusion that is especially visible in the DALL·E images, which often have uninterpretable words overlaying on some background images. Only when language and visual communication play an important role (as in the Tower of Babel story), we may see a link between the text prompt and the text as part of the generated image. 
  In future work, some textual pre-processing could be performed to provide the generators with more detailed instructions and less information that can be difficult for visualization, such as conversations.

\section{Conclusion and Future Work}
\label{sec:conclusion}

This paper presented an interdisciplinary approach to studying AI-generated biblical art. We proposed a systematic evaluation of the images generated with biblical text as their prompts. For RQ1 (``How can we systematically generate biblical images using text-to-image generators?''), we selected biblical text as prompts and generated, using various generators, a large dataset with over 7K images.  RQ2 (``How can we evaluate the biblical images generated?'') was tackled with the help of different neural network-based image assessment models. We chose five paintings for each prompt as references for evaluation. We proposed measures for the assessment of accuracy. Our analysis of numbers on humans, gender, and age answered RQ2a (``What is the accuracy of the persons and objects in the generated images regarding their biblical context?''). As for RQ2b ``How can we compare the sentimental values of the generated images?''), we employed two models to obtain the sentimental values. For RQ2c (``What features can be analyzed for the generated images concerning religion and aesthetics?''), we performed some manual analysis and provide an analysis regarding different aspects of religion and aesthetics.

\textbf{Based on the evaluation using our measures, Midjourney generates illustration-like images that are most similar to selected human artworks in our base with sophisticated details and some reasonable understanding of the given religious context.} We noticed limited diversity in generated images. Some future work could be focused on overcoming this by changing the parameters, alternating the input prompt, etc.  The variants of Stable Diffusion perform more similar among themselves than dissimilar. Many images present some understanding of the religious context with more diversity in styles. In contrast, DALL-E is the generator that captures the religious context the least with images generated most different from selected human art. Finally, we discussed the limits of our approach with reflection on the evaluation results and the features of the AI generators.

There are many issues that require further investigation. There are objects other than humans in the selected biblical text including altar, wood, knife, as well as spiritual beings. Some images by Midjourney and Stable Diffusion can generate such objects based on our manual assessment but not always perfectly. How the generators recognize typical scenes such as the Last Supper and the Tower of Babel require further investigation. How we can further manipulate and improve the prompts to improve the accuracy of generated images deserves further study. Some prompts had truncated text while loaded to generators due to the maximum size of the prompt. We could further assess images produced based on different truncated versions. We noticed that in some images, some males were recognized as females. This calls for a benchmark on the models' performance using generated images. Otherwise, these errors can have an impact on the evaluation result. It remains to be studied how we could use paintings beyond the Renaissance and Baroque artworks to evaluate the performance.  Moreover, the workflow could be extended to incorporate additional Machine Learning and image processing models that classify landscape, facial emotion, and weather, etc. This could contribute to future research on how generated images vary in context, accuracy, art style, theme, and other interpretive features. In addition, some existing deep learning models can be used to evaluate memorability \cite{cetinic2019deep}.

In this paper, the King James Version (KJV) was used. In the future, the effect of the Bible translation used for the prompts could be analyzed. Even in cases where the generators did not get the narrative details of a prompt, they seemed to have captured its religious character by inserting halos or crosses (cf. Sections \ref{sec:method-DALLE} and \ref{sec:StableDiffusion}). Is this related to the archaic style of the King James Version that we used? What happens if we use a modern translation such as the Good News Bible? 

The generated images, the code, and the corresponding assessment results are open source. They can be used for future research for the comparison of benchmarks of the performance of generators on other topics in art, theology, and computer science.

Finally, although our dataset is published as an open source, finding an image with specific features can be difficult due to its large size. A possible future work is a platform for retrieving images with certain features (e.g. ``six males and two females'', ``dark background'', or a given sentimental range) according to the scores described in Section \ref{sec:evaluation}. As A.I. rapidly advances in its ability to create images, such a platform could serve as a repository for storing and comparing images produced by other recent and future A.I. tools.


\subsection*{Acknowledgement}

The authors appreciate the help of colleagues of the ETCBC and the Network Institute, especially Ivano Malavolta. The authors used TeXGPT in combination with Writefull to rephrase some phrases. No sentence or paragraph was generated by TeXGPT or any other A.I. tool.

\appendix

\section{Prompts used for the generation of images}
\label{appendix:prompt}

The following are the original texts selected from the King James Version (KJV) of the Bible. See Section \ref{sec:ImageGeneration} for the truncation of prompts 2 and 4 using the NLTK Library. Each of them served as a prompt without additional instructions, such as ``Generate a picture of the biblical Tower of Babel'' or ``of Jesus at the Last Supper''.

\subsubsection*{Prompt 1}

`therefore the Lord God sent [Adam and Eve] forth from the garden of Eden, to till the ground from which he was taken. He drove out the man; and at the east of the garden of Eden he placed the cherubim, and a flaming sword which turned every way, to guard the way to the Tree of life.

\subsubsection*{Prompt 2}
Now the whole earth had one language and few words. And as men migrated from the east, they found a plain in the land of Shinar and settled there. And they said to one another, ``Come, let us make bricks, and burn them thoroughly.'' And they had brick for stone, and bitumen for mortar. Then they said, ``Come, let us build ourselves a city, and a tower with its top in the heavens, and let us make a name for ourselves, lest we be scattered abroad upon the face of the whole earth.'' And the Lord came down to see the city and the tower, which the sons of men had built. And the Lord said, ``Behold, they are one people, and they have all one language; and this is only the beginning of what they will do; and nothing that they propose to do will now be impossible for them. Come, let us go down, and there confuse their language, that they may not understand one another’s speech.'' So the Lord scattered them abroad from there over the face of all the earth, and they left off building the city. Therefore its name was called Babel, because there the Lord confused[a] the language of all the earth; and from there the Lord scattered them abroad over the face of all the earth.

\subsubsection*{Prompt 3}
When they came to the place of which God had told him, Abraham built an altar there, and laid the wood in order, and bound Isaac his son, and laid him on the altar, upon the wood. Then Abraham put forth his hand and took the knife to slay his son. But the angel of the Lord called to him from heaven, and said, ``Abraham, Abraham!'' And he said, ``Here am I.''
He said, ``Do not lay your hand on the lad or do anything to him; for now I know that you fear God, seeing you have not withheld your son, your only son, from me.'' And Abraham lifted up his eyes and looked, and behold, behind him was a ram, caught in a thicket by his horns; and Abraham went and took the ram, and offered it up as a burnt offering instead of his son. So Abraham called the name of that place The Lord will provide;[a] as it is said to this day, ``On the mount of the Lord it shall be provided.''

\subsubsection*{Prompt 4}
And when it was evening he came with the twelve. And as they were at the table eating, Jesus said, Jesus said, ``Truly, I say to you, one of you will betray me, one who is eating with me.'' They began to be sorrowful, and to say to him one after another, ``Is it I?'' He said to them, ``It is one of the twelve, one who is dipping bread into the dish with me. For the Son of man goes as it is written of him, but woe to that man by whom the Son of man is betrayed! It would have been better for that man if he had not been born.'' And as they were eating, he took bread, and blessed, and broke it, and gave it to them, and said, ``Take; this is my body.'' And he took a cup, and when he had given thanks he got gave it to them, and they all drank of it. And he said to them, ``This is my blood of the[b] covenant, which is poured out for many. Truly, I say to you, I shall not drink again of the fruit of the vine until that day when I drink it new in the kingdom of God.''

\subsubsection*{Prompt 5}
Now the daughter of Pharaoh came down to bathe at the river, and her maidens walked beside the river; she saw the basket among the reeds and sent her maid to fetch it. When she opened it she saw the child; and lo, the babe was crying. She took pity on him and said, ``This is one of the Hebrews’ children.'' Then his sister said to Pharaoh’s daughter, ``Shall I go and call you a nurse from the Hebrew women to nurse the child for you?'' And Pharaoh’s daughter said to her, ``Go.'' So the girl went and called the child's mother. And Pharaoh’s daughter said to her, ``Take this child away, and nurse him for me, and I will give you your wages.'' So the woman took the child and nursed him.

\section{Use Cases: VR Exhibition}

\label{sec:vr}

\begin{figure}[!ht]
\centering
\minipage{0.95\textwidth}
  \includegraphics[width=\linewidth]{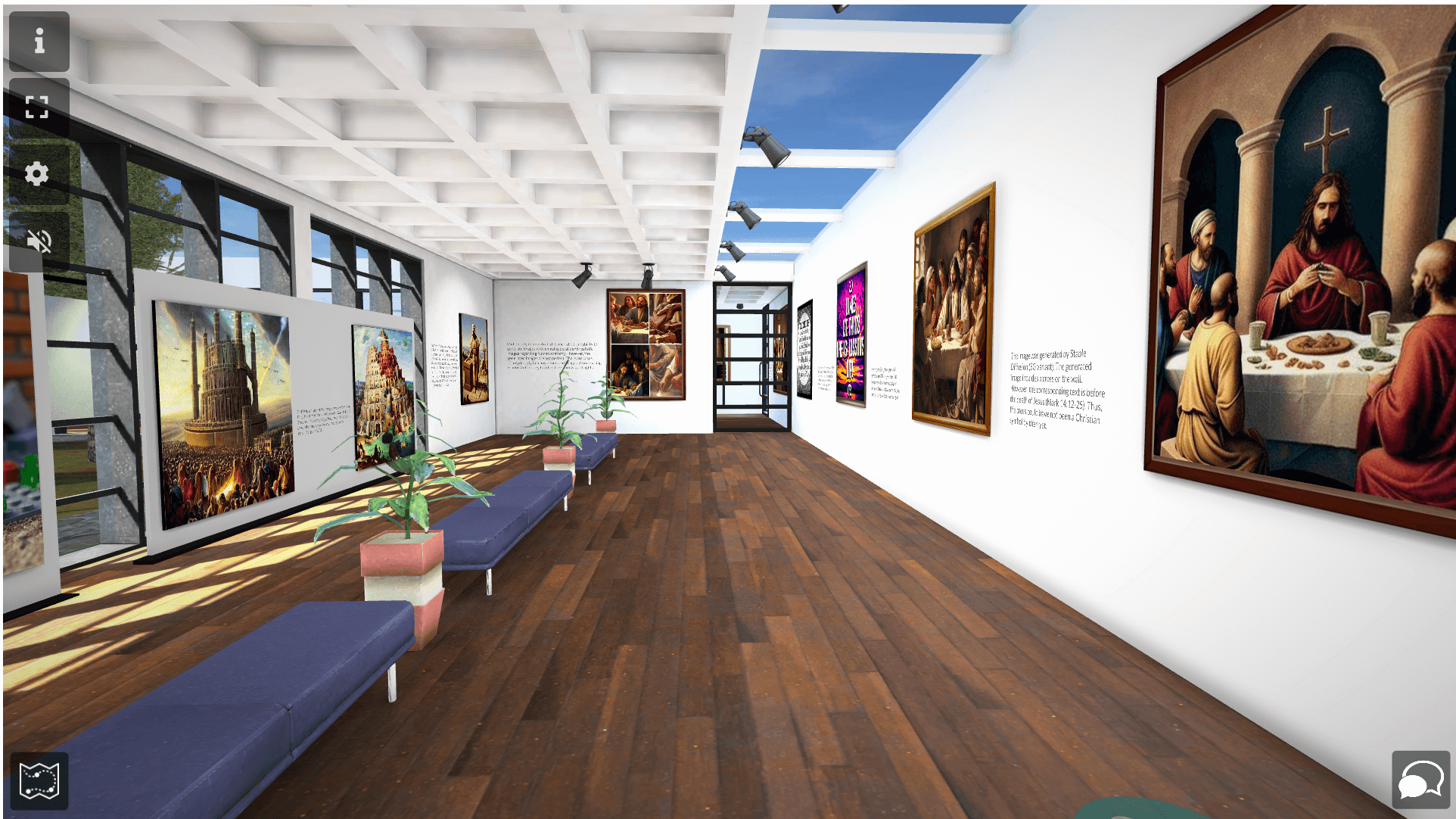}
  \caption{A view from the entrance of the VR gallery}\label{fig:vr1}
  \label{fig:vr_view1}
\endminipage\hfill
\end{figure}

\begin{figure}[!ht]
\centering
\minipage{0.95\textwidth}
  \includegraphics[width=\linewidth]{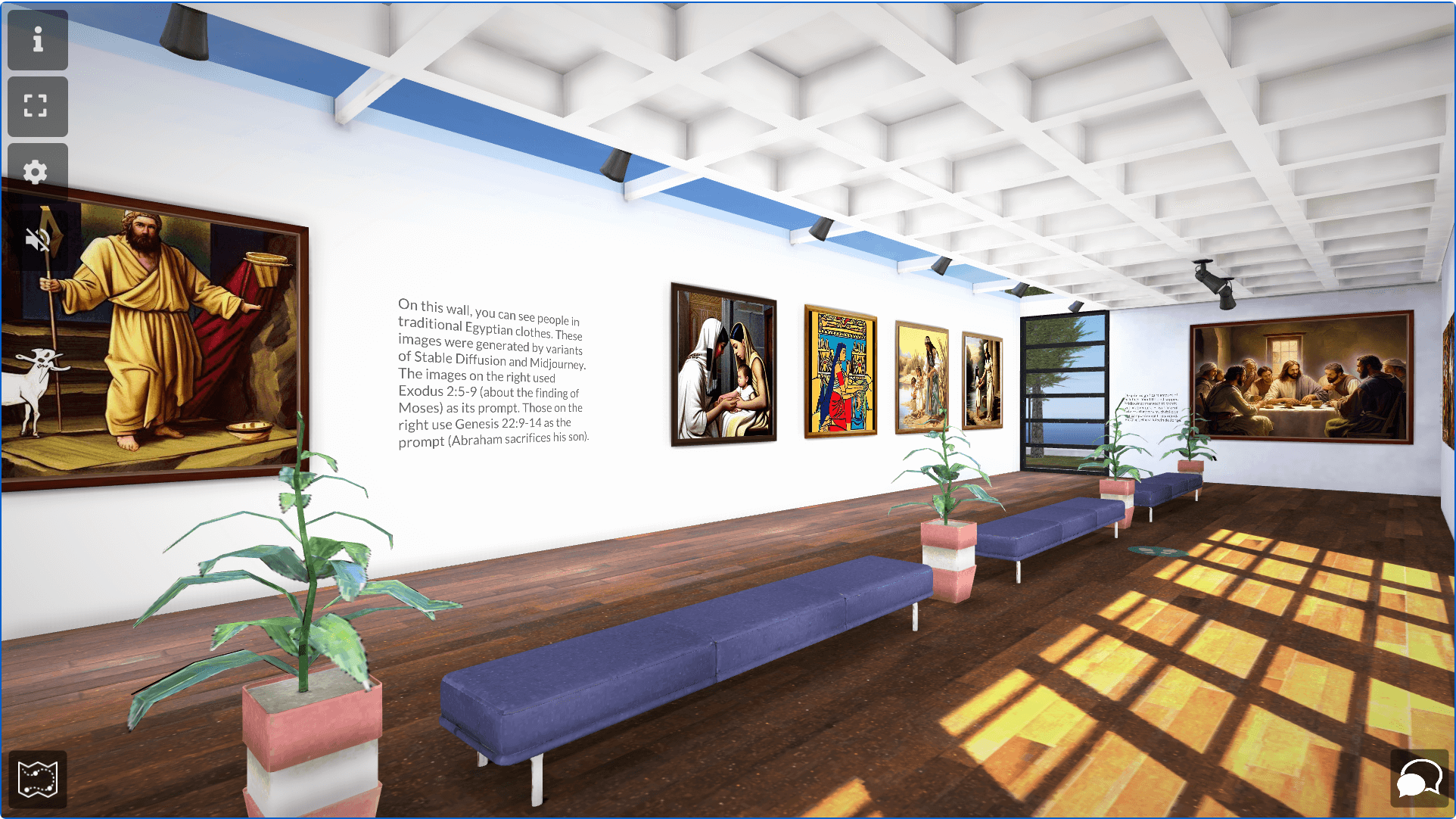}
  \caption{A view next to the exit of the VR gallery}\label{fig:vr_view2}
\endminipage\hfill
\end{figure}

In this appendix, we provide a use case of a virtual reality exhibition\footnote{\url{https://shuai.ai/art/seeing}}. The exhibition uses the ArtSteps platform\footnote{\url{https://www.artsteps.com/}}, which enables the customization of a semi-defined virtual reality space. Figure \ref{fig:vr_view1} shows the entrance of the VR gallery while Figure \ref{fig:vr_view2} shows the view near the exit of the gallery. Next to the images are some explanatory text about that generation of the images together with some analysis. The exhibition can also be viewed on a mobile phone using the ArtSteps app with a VR headset.

\section{Exceptional Images Generated}
\begin{figure}[!ht]

\minipage{0.45\textwidth}
\includegraphics[width=\linewidth]{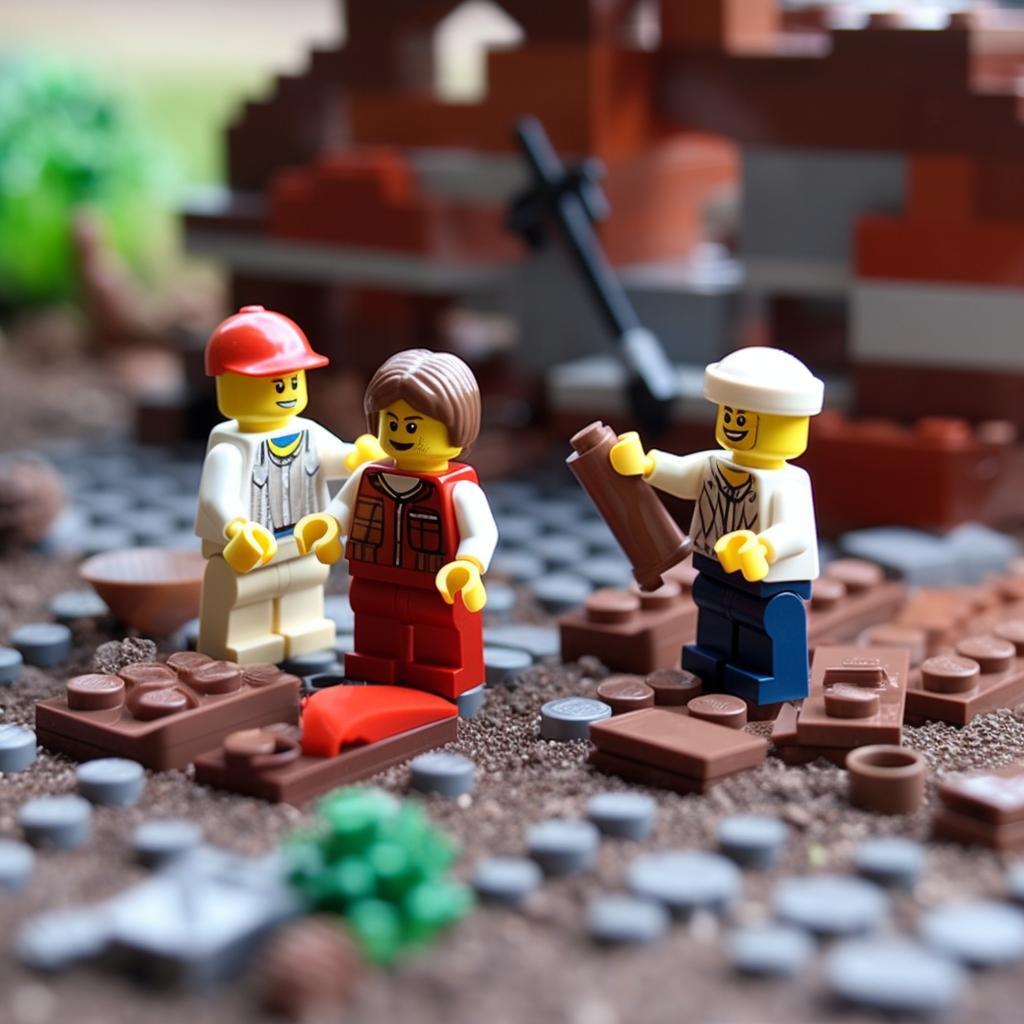}
   \caption{A generated image by Midjourney using prompt 2 (the Babel Tower). Instead of humans, there are three LEGO minifigures working on a construction site.}\label{fig:lego1}
\endminipage\hfill
\minipage{0.45\textwidth}
\includegraphics[width=\linewidth]{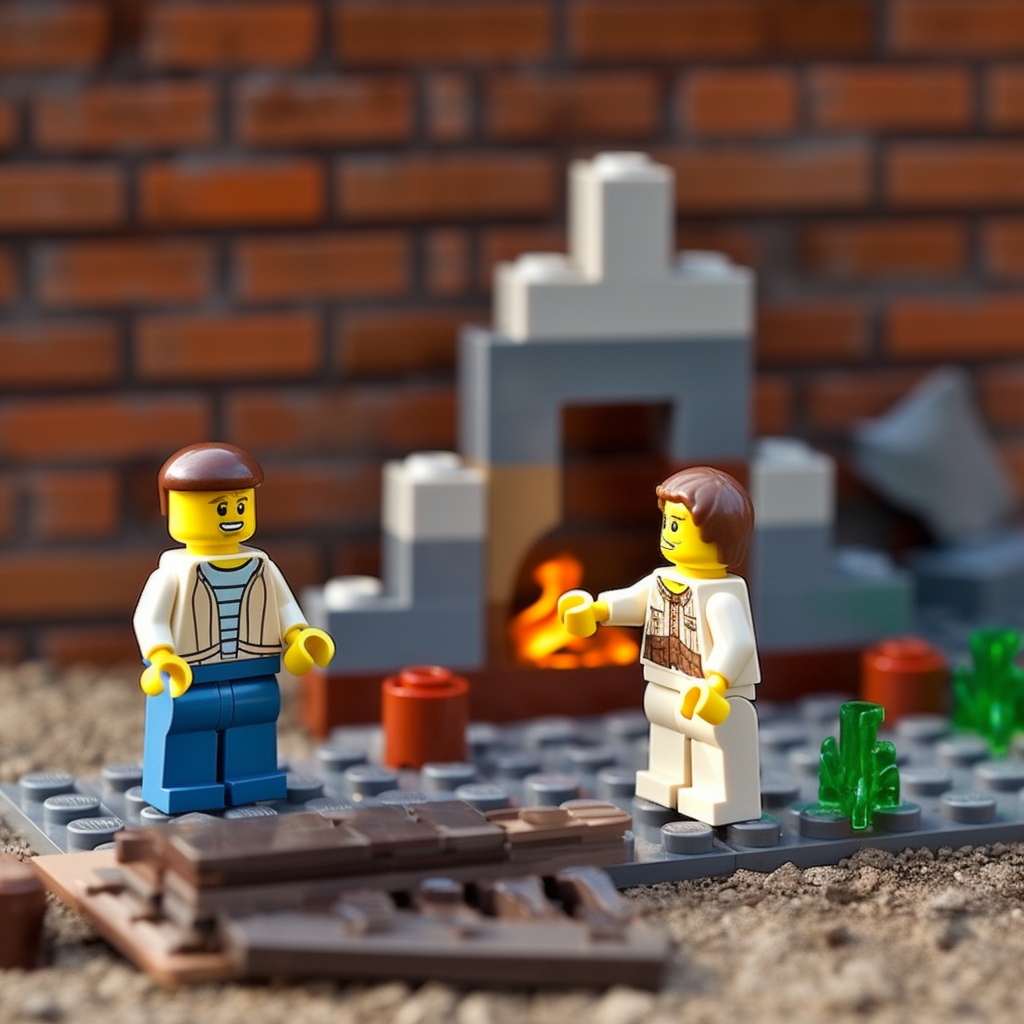}
   \caption{A generated image by Midjourney using prompt 2 (the Babel Tower). Instead of humans, there are two LEGO minifigures working on a construction site.}\label{fig:lego2}
\endminipage
\end{figure}

{\color{black}
Here we discuss some exceptional images that were found unexpectedly in while doing manual analysis. Despite the observation that Midjourney generates consistently illustration-like images such as Figure \ref{fig:last4} and \ref{fig:last5}, it was not expected to encounter Figure \ref{fig:lego1} and \ref{fig:lego2}, two exceptional images. These images do not show any human character. Instead, there are some LEGO minifigures on a construction site. Upon further examination, it was revealed that there are some videos on YouTube that use LEGO minifigures to explain Bible stories\footnote{Such as \url{https://www.youtube.com/watch?v=lgSkv68hk7A} and \url{https://www.youtube.com/watch?v=zTlfjz30QSk}.}. This shows some degree of creativity, but triggers the question if Midjourney used images on YouTube as a part of their training data and some discussion if such representation is serious enough for this religious setting. }

\newpage
\bibliographystyle{splncs04}
\bibliography{main}

\end{document}